\documentclass[aip,jcp,preprint,superscriptaddress]{revtex4-1}

\usepackage{amsmath}
\usepackage{graphicx}
\usepackage{color}

\begin{document}

\title{Unidirectional hopping transport of interacting particles
 on a finite chain}

\author{Mario Einax}
\email{mario.einax@tu-ilmenau.de}
\homepage{http://www.tu-ilmenau.de/theophys2}
\affiliation{Institut f\"ur Physik, Technische Universit\"at
Ilmenau, 98684 Ilmenau, Germany} \affiliation{School of Chemistry,
Tel Aviv University, Tel Aviv 69978, Israel}

\author{Gemma Solomon}
\affiliation{Department of Chemistry, Northwestern University,
Evanston, Illinois 60208, USA}

\author{Wolfgang Dieterich}
\affiliation{Fachbereich Physik, Universit\"at Konstanz, 78457
Konstanz, Germany}

\author{Abraham Nitzan}
\affiliation{School of Chemistry, Tel Aviv University, Tel Aviv
69978, Israel}

\date{\today}

\begin{abstract}
Particle transport through an open, discrete 1-D channel against
a mechanical or chemical bias is analyzed within a master equation
approach. The channel, externally driven by time dependent site energies,
allows multiple occupation due to the coupling to reservoirs.
Performance criteria and optimization of active transport in a two-site channel
are discussed as a function of reservoir chemical potentials,
the load potential, interparticle interaction strength, driving mode
and driving period.
Our results, derived from exact rate equations, are used in addition
to test a previously developed time-dependent
density functional theory, suggesting a wider applicability
of that method in investigations of many particle systems far
from equilibrium.
\end{abstract}


\maketitle

\section{Introduction}
\label{Introduction} Random motion of a classical particle in a
potential that breaks spatial inversion symmetry and fluctuates in
time generally leads to unidirectional flow. On the molecular
level, many important processes in biology and nanotechnology rely
on this mechanism. Biological motors produce mechanical work from
metabolic energy in order to effect intracellular transport, or
self-propulsion of bacteria through rotatory flagellar motion.
\cite{Juelicher97} Another process ubiquitous in any living
organism is active transport of molecules or ions against a
chemical potential gradient. Such molecular or ionic pumps
generally consist of a specific channel across a cell membrane.
The internal binding sites of the channel are linked to
conformational fluctuations, stimulated by metabolic energy or by
light. \cite{Lauger91,Hille01,Muneyuki00,Kaila08} Similar
processes are known for artificial nanopores, with potential
applications in molecule or ion separating devices.
\cite{Rousselet94,Siwy02} In the realm of quantum transport, a net
electron drift under an applied ac driving signal can be generated
by various mechanisms, which may become analogous to classical
unidirectional transport when dissipation is included.
\cite{Galperin05,Altshuler99}

Our goals in this paper are twofold. First, we set up a model that
describes unidirectional transport along a finite open system,
represented by a non-symmetric discrete chain with time-dependent
driving. This model emphasizes {(i)} coupling of the two chain
ends to reservoirs and {(ii)} interaction effects between
transported particles that occupy the chain. These combined
features distinguish our study from most works on Brownian
ratchets, reviewed in \cite{Reimann02,Haenggi09}. Indeed, both
interacting Brownian motors
\cite{Derenyi95,Savelev05,Reimann99,Slanina08, Slanina09} and ion
pumping mechanisms that involve multiple occupation of the
associated channel structure \cite{Muneyuki00} have become
important subjects of research in nonequilibrium statistical
physics and biophysics. The model we examine allows us to study
production of both mechanical and chemical work. In what follows
we refer to a machine working against a mechanical force as
"motor" and to that working against a chemical load as a
``chemical pump''. \cite{Juelicher97} Accordingly, attention will
be focused on the system performance in both modes of operation
and its dependence on the driving characteristics and on
fundamental input parameters such as chemical potentials of the
reservoirs, load potential, and interaction strengths. Although
our model is certainly far from describing realistic systems, we
argue below that at a qualitative level, the effects of
concentration and interaction evaluated here should have rather
general validity.

Our second goal is to provide a test of time-dependent density
functional theory (TDFT) \cite{Reinel96,Kessler02,Gouyet03} when
applied to far from equilibrium dynamics under time-dependent
driving signals. \cite{note1} This method can be regarded as a
version of dynamic mean-field theory, constructed such that it can
account for exact static properties. Its accuracy in predicting
transport properties as envisaged here will be assessed by a
comparison with numerical solutions of the underlying master
equation for short chains containing four sites. The numerical
effort needed for such solutions increases exponentially with
system size (here expressed by the number of sites between the
left and right reservoirs). For this reason, establishing the
validity of an approximate solution is important for future
applications.

After defining our model in Sec.~\ref{sec:Definition}, we present
in Sec.~\ref{sec:4sites} a minimal description in terms of a
$4$-site model. Section~\ref{sec:TDFT} introduces the TDFT method,
while Sec.~\ref{sec:results} examines its accuracy in comparison
to ``exact'' numerical solutions. Also in Sec.~\ref{sec:results}
we present results concerning the performance of the system
studied, focusing on the efficiency of its operation either as a
motor working against a mechanical or electrical load, or as a
chemical pump acting against a chemical bias. Section
\ref{sec:Summary} concludes.

\section{Model}
\label{sec:Definition} Our system is a $1$-D ``Fermionic'' lattice
gas with sites $l=1,...,M$, time-dependent site energies
$\varepsilon_{l} (t)$, nearest neighbor hopping, and a nearest
neighbor interaction $V$. By considering a ``Fermionic'' lattice
gas, equivalent to local hard core repulsions, effects of
saturation of site occupations are automatically included. The
sites $l=1$ and $M$ can exchange particles with left and right
reservoirs, $L$ and $R$, that are sometimes represented below by
indices $0$ and $M+1$, respectively.
We assume that these reservoirs exchange particles with the system
with characteristic specified rates, but we disregard interactions
between particles in the systems and those in the reservoirs, i.e.
$V_{0,1}=V_{M,M+1}=0$. By definition, reservoir particles are
always in equilibrium and have fixed mean occupations,
\begin{align}\label{eq:pLpR}
  p_{J} &= \left(e^{-\beta\mu_{J}}+1\right)^{-1}\, ,  \quad J=L,R
\end{align}
where $\mu_L$, $\mu_R$ are the respective reservoir chemical
potentials, and $\beta=1/k_BT$. Rate equations for the averaged
site occupations $p_l(t);\, l=1,...,M;$ are given by
\begin{align}
\label{eq:p_l}
  \frac{dp_{l}}{dt} &= \left\langle
  j_{l-1,l}\right\rangle_t - \left\langle
  j_{l,l+1}\right\rangle_t\,,
\end{align}
where $\left\langle j_{l,l+1}\right\rangle_t$ for $l=0,...,M$
denotes the net average current from site $l$ to $l+1$, to be
derived from the underlying master equation. Note that
$\left\langle
  j_{0,1}\right\rangle_t \equiv \left\langle j_{L,1}\right\rangle_t$
and $\left\langle j_{M,M+1}\right\rangle_t \equiv \left\langle
  j_{M,R}\right\rangle_t$ are currents from the left and to the right
reservoir. In order to proceed, we need to specify the rates for
configurational transitions consistent with the detailed balance
condition. We adopt here symmetric rates
\begin{align}
\label{eq:rate}
  w_{i,f} \propto \exp\left[\beta(E_i -
  E_f)/2\right] \, ,
\end{align}
where $E_i$ and $E_f$ is the total energy in the initial and final
state, respectively. In particular, `bare' transition rates for
elementary hops from $l$ to $l\pm 1$, that govern the single particle
dynamics in the dilute limit, are given by $k_{l,l \pm 1} (t)=\nu_{l,l
  \pm 1} \exp\left[\beta( \varepsilon_l(t) - \varepsilon_{l\pm
    1} (t))/2\right]$. Here $\nu_{l,l\pm 1}$ are frequency factors,
which for simplicity are assumed independent of $l$ for
$l=1,\cdots M-1$ and represent the bulk frequency $\nu_B$, while
at the system boundary we distinguish
$\nu_{L}=\nu_{0,1}=\nu_{1,0}$ and
$\nu_{R}=\nu_{M,M+1}=\nu_{M+1,M}$ from $\nu_B$. Setting
$(\nu_B)^{-1}=1$ and $\beta=1$ defines our units of time and
energy.

Certain special cases of this model deserve special attention. For
static site energies $\varepsilon_l$ it describes aspects of
passive transport, for example, through membrane channels.
\cite{Kolomeisky07,Berezhkovskii/etal:2008,Zilman/etal:2009} Very
recently, the nonlinear dc response and rectification in a single
particle hopping system coupled to reservoirs were examined,
including disorder effects. \cite{Einax09} Assuming $p_L \not=
p_R$ but taking $\varepsilon_l$ independent of both time and
space, one recovers a generalized asymmetric simple exclusion
process (ASEP) model \cite{Dierl/etal:2010} that contains a
nearest neighbor coupling $V$ (for reviews of the hard-core
repulsion ASEP as well as the totally asymmetric simple exclusion
process (TASEP) model, see
Refs.~\onlinecite{Derrida:1998,Schuetz:2001} and references
therein).

Several driving modes, both stochastic and deterministic, that
lead to unidirectional transport, were proposed in the literature.
\cite{Reimann02,Haenggi09} In this work we assume for simplicity a
sinusoidal time dependence with frequency $\omega=2\,\pi/\tau$,
which should allow us to study transport efficiencies as a
function of the typical time scale set by the modulation period
$\tau$ that characterizes the driving forces. With regard to the
spatial asymmetry, common models are: \cite{Haenggi09}
\begin{itemize}
  \item[{(a)}] Peristaltic or travelling wave-like behavior, where
    potential minima move in one direction, thereby dragging particles
    with them.
  \item[{(b)}] Sawtooth-like potential with oscillating amplitude
    (``flashing ratchet''). In such systems particles are driven in
    the direction against the steeper potential slope.
  \item[{(c)}] Constant ($l$-independent) ac-force (``rocking ratchet'')
    superimposing a non-symmetric static potential. Unidirectional
    transport directly results from steady state rectification
    properties of the static potential, as can be seen by considering
    the adiabatic limit $\omega \rightarrow 0$.
\end{itemize}
Unidirectional transport in the latter model results from its DC
rectification properties determined by the static potential, as
can be seen by considering the adiabatic limit $\omega \rightarrow
0$. The present work focuses on the first two mechanisms, (a) and
(b). The model introduced above applies to multiply occupied
channels driven by time dependent mechanical forces, and working as
motors or as chemical pumps. As specific examples we consider the
following situations:
\begin{itemize}
 \item[{(i)}] \emph{Motor action}:\,
    A constant load $F$ is applied that changes the site energies
    relative to their intrinsic values $(\varepsilon_l)_{F=0}$,
    \begin{align}
    \label{eq:epsilon}
        \varepsilon_l&=(\varepsilon_l)_{F=0}+\frac{Fl}{(M+1)};
        \,\quad \,l=0,...,M+1.
    \end{align}
    By this, the left and right reservoirs acquire a potential energy
    difference $\varepsilon_R - \varepsilon_L = F$,
    while their chemical potentials are taken equal, $p_R=p_L$.
 \item[{(ii)}] \emph{Chemical pump}:\,
    In a pure chemical pump the chemical potentials in the two reservoirs are
    different, e.g., $\mu_R >\mu_L$ ($p_R > p_L$),
    while the mechanical load F
    vanishes. Mathematically this case differs from the model
    (\ref{eq:epsilon}) by the fact that, unlike the linear potential change in
    (\ref{eq:epsilon}), the chemical potential difference $\mu_R-\mu_L$
    will in general not give rise to a linear distribution of local
    chemical potential changes along the channel.
\end{itemize}

In both cases we can discuss different measures of machine
performance and optimization schemes. First the output of the
machine operation can be obtained from the average current
\begin{align}
    J_{{\rm av}} &=
    \frac{1}{\tau}\int_{0}^{\tau}dt \frac{1}{M+1}
    \sum_{l=0}^{M}\left\langle
    j_{l,l+1}\right\rangle _{t} \label{eq:j_av}
\end{align}
where, at steady state, the average (over realizations and over a
modulation period $\tau$) currents between neighboring positions
do not depend on position. The useful work output is then
\begin{align}
\label{eq:W_out_mech}
      \overline{W}_{{\rm out}} &= J_{{\rm av}} F \, .
\end{align}
for the pure motor action, and
\begin{align}
\label{eq:W_out_chem}
      \overline{W}_{{\rm out}} &= J_{{\rm av}} (\mu_R -\mu_L) \, .
\end{align}
for the pure chemical pump. Note that combined effects of
mechanical and chemical biases can be considered, in which case
$\overline{W}_{{\rm out}} = J_{{\rm av}} (\mu_R^e -\mu_L^e)$ is
the sum of (\ref{eq:W_out_mech}) and (\ref{eq:W_out_chem}), the
electrochemical potentials $\mu_{L,R}^e$ being defined according
to (\ref{eq:electrochem_pot}). In all cases the work input can be
calculated from
\begin{align}
\label{eq:W_in}
      \overline{W}_{{\rm in}} &=
      \frac{1}{\tau}\int_{0}^{\tau}dt \sum_{l=1}^{M} \frac{d
      \varepsilon_l(t)}{dt} p_l (t)
\end{align}
This yields the conventional "efficiency" \cite{Sumithra01}
\begin{align}
       \label{eq:efficiency} \eta &= \frac{\overline{W}_{{\rm
       out}}}{\overline{W}_{{\rm in}}} \, .
\end{align}
Aiming for maximum efficiency is one criterion for optimization.
Alternatively, irrespective of the amount of input energy, one can
ask for the maximum F, where $J_{{\rm av}}$ changes sign
(``reversal potential''), or for
the maximum current $J_{{\rm av}}>0$ against the load $F>0$,
corresponding to the maximum rate of transfer of particles from
$L$ to $R$.

\section{Four-site model}\label{sec:4sites}
Setting $M=2$, we have a $4$-site system consisting of a channel
with sites $l=1,\,\,2$ in contact with boundary sites $L$ and $R$.
In this minimal ratchet model open to reservoirs the system is
driven by modulating the site energies $\varepsilon_1(t)$ and
$\varepsilon_2(t)$. An incipient peristaltic modulation of site
energies can be realized by a phase lag in the oscillation of
$\varepsilon_2(t)$ relative to $\varepsilon_1(t)$,
\begin{align}
\varepsilon_{L} &= 0
\label{eq:peristaltic} \\
  \varepsilon_1(t) &= \varepsilon_1^{(0)}+A\left[ 1+\sin (\omega t)
  \right]+F/3
\label{eq:peristaltic_1}\\
  \varepsilon_2(t) &= \varepsilon_2^{(0)}+A\left[ 1+\sin (\omega
  t-\pi/2) \right]+2F/3
\label{eq:peristaltic_2}\\
  \varepsilon_{R} &= F \, . \label{eq:mechanical_gradient}
\end{align}
Here, the energies $\varepsilon_l^{(0)}$ represent a constant
energy shift of the channel's interior relative to the reservoirs.
On the other hand, the in-phase oscillation
where Eqs. (\ref{eq:peristaltic_1}) and (\ref{eq:peristaltic_2})
are replaced by
\begin{align}\label{eq:flashing}
  \varepsilon_1(t)&= \varepsilon_1^{(0)}+2A \left[1+ \sin(\omega t) \right] + F/3 \\
  \varepsilon_2(t)&= \varepsilon_2^{(0)}+A \left[1+ \sin(\omega t)
  \right] + 2F/3
  \label{eq:flashing_e2}
\end{align}
(keeping (\ref{eq:peristaltic}) and
(\ref{eq:mechanical_gradient})), corresponds to a ``flashing''
ratchet of the type of a discrete ``sawtooth'' potential. Both of
these driving modes favor a current to the right.

Unidirectional flow induced by such driving schemes can be most
simply investigated in the independent particle model where the
average site occupations  $p_1(t)$ and $p_2(t)$  evolve according
to the linear rate equations
\begin{align}
\label{eq:p1dt} \frac{d p_1}{ dt} &= k_{L,1}(t) p_L - k_{1,L}(t) p_1 +
k_{2,1}(t) p_2 - k_{1,2}(t) p_1\\
\label{eq:p2dt} \frac{d p_2}{ dt} &= k_{R,2}(t) p_R - k_{2,R}(t)
p_2 - k_{2,1}(t) p_2 + k_{1,2}(t) p_1
\end{align}
with prescribed populations $p_{L,R}$ at the boundary sites. A
more realistic model should take into account interparticle
interactions. Here we consider both hard core (site exclusion) and
nearest neighbor interaction, denoted $V$.

The following section (Sec.~\ref{sec:TDFT}) describes an
approximate approach to the kinetics of such models based on the
(classical) time-dependent density functional theory. An exact
approach, feasible for the present small system is based on rates
equations written in the system states representation. A system
state $(n_1,n_2)$ is defined in terms of the occupations of sites
$l=1$ and $l=2$. Obviously our system is fully characterized by
the four states $(0,0)$, $(1,0)$, $(0,1)$ and $(1,1)$, with the
corresponding energies
\begin{align}
\label{eq:states_energy_10}
  E_{10}(t)&=\varepsilon_1 (t)\\
  E_{01}(t)&=\varepsilon_2 (t)\\
\label{eq:states_energy_11}
  E_{11}(t)&=\varepsilon_1 (t)+\varepsilon_2 (t)+V.
\end{align}
$E_{00}$ may conveniently be set to $0$. We assume that
transitions between these states proceed only by single particle
steps, so that no direct transition takes place between states
$(0,0)$ and $(1,1)$. It should be emphasized that in using these
energies to determine rates one needs to take into account the
change of energy in the reservoir. Thus, the total energy change
$\Delta E$ for the transition $(0,0)\rightarrow (1,0)$ is $E_{10}
-\varepsilon_L = \varepsilon_1-\varepsilon_L$ and $\Delta E$ for
the transition $(1,0)\rightarrow (1,1)$ is
$E_{11}-E_{10}-\varepsilon_R = \varepsilon_2+V-\varepsilon_R$. The
kinetic equations for the average population of these system
states are
\begin{align}
\label{eq:dp_10}
  \frac{dP_{10}(t)}{dt}&=K_{00,10}(t) P_{00}(t) + K_{01,10}(t) P_{01}(t) + K_{11,10}(t) P_{11}(t) \\
  & - \left( K_{10,00}(t)+K_{10,01}(t)+K_{10,11}(t)\right)
  P_{10}(t) \nonumber \\
\label{eq:dp_01}
  \frac{dP_{01}(t)}{dt}&=K_{00,01}(t) P_{00}(t) + K_{10,01}(t) P_{10}(t) + K_{11,01}(t) P_{11}(t) \\
  & - \left( K_{01,00}(t)+K_{01,10}(t)+K_{01,11}(t)\right) P_{01}(t) \nonumber \\
\label{eq:dp_11}
  \frac{dP_{11}(t)}{dt}&=K_{10,11}(t) P_{10}(t) + K_{01,11}(t)
  P_{01}(t)-\left( K_{11,10}(t)+K_{11,01}(t)\right) P_{11}(t)
\end{align} and normalization implies that
\begin{align}
\label{eq:norm}
  P_{00}=1-(P_{10}+P_{01}+P_{11}) \, .
\end{align}
Using (\ref{eq:norm}) in (\ref{eq:dp_10})-(\ref{eq:dp_11}) yields
three inhomogeneous equations with rates that are readily obtained
from (\ref{eq:rate}) and
(\ref{eq:states_energy_10})-(\ref{eq:states_energy_11}). Clearly,
$K_{01,10}(t)=k_{2,1}(t)$ and $K_{10,01}(t)=k_{1,2}(t)$. The
remaining ones involve the bath densities, for example,
\begin{align}
\label{eq:gamma_states}
K_{10,00}(t)&= \nu_L \exp[\beta (\varepsilon_1(t)-\varepsilon_L)/2] (1-p_L)\\
K_{00,10}(t)&= \nu_L \exp[-\beta (\varepsilon_1(t)-\varepsilon_L)/2] p_L
\end{align}
Once the solutions to Eqs.~(\ref{eq:dp_10})-(\ref{eq:norm}) have
been found, the average populations of individual  sites are
obtained from:
\begin{align}
\label{eq:p2_p3}
  p_1 (t)&=P_{10}(t)+P_{11}(t)\\
  p_2 (t)&=P_{01}(t)+P_{11}(t).
\end{align}
and the currents $J_L(t)\equiv \left\langle
j_{L,1}\right\rangle_t$; $J_R(t)\equiv  \left\langle
j_{2,R}\right\rangle_t$ can be calculated from, e.g.
\begin{align}
\label{eq:current_state_model}
   J_L(t) &=  K_{00,10} P_{00} -
   K_{10,00} P_{10} + K_{01,11} P_{01} - K_{11,01} P_{11}, \\
   J_R(t) &=  K_{01,00} P_{01} -
   K_{00,01} P_{00} + K_{11,10} P_{11} - K_{10,11} P_{10}.
\end{align}

Note that results based on Eqs.~(\ref{eq:dp_10})-(\ref{eq:norm})
differ from those of the independent particle model,
Eqs.~(\ref{eq:p1dt})-(\ref{eq:p2dt}), even in the limit $V=0$,
because unlike the latter they incorporate hard core interactions.
The behavior based on Eqs.~(\ref{eq:p1dt})-(\ref{eq:p2dt}) is
expected only in the highly dilute limit when particle encounters
on the same site are negligible.

\section{Time-dependent density functional theory (TDFT)}
\label{sec:TDFT} TDFT is a local equilibrium approximation, in
which the non-equilibrium character of the distribution function
is manifested in the space- and time-dependent local fields acting
on the single particle density. Given a distribution function of
this type, density functional theory assumes that correlators
determining the currents $\left\langle j_{l,l+1}\right\rangle_t$
in Eq.~(\ref{eq:p_l}) are functionals of the single particle
density, and moreover, that this functional dependence is the same
as in the equilibrium case. This allows us to express the mean
currents in (\ref{eq:p_l}) in terms of the densities $p_l(t)$ and
therefore to arrive at a closed system of nonlinear rate
equations. In the present problem this latter step can be carried
through exactly because the free energy functional for
$1$-dimensional lattice gases with short range interactions is
known. \cite{Percus94,Buschle00}

The rate equations (\ref{eq:p_l}) for the averaged site
occupations $p_l(t);\,\, l=1,\,2;$ in the $4$-site model reduce to
\begin{align}
\label{eq:p_1}
 \frac{dp_{1}}{dt} &= \left\langle
    j_{L,1}\right\rangle_t -
    \left\langle j_{1,2}\right\rangle_t \, ,\\
 \frac{dp_{2}}{dt} &= \left\langle j_{1,2}\right\rangle_t
    -\left\langle j_{2,R}\right\rangle_t \, . \label{eq:p_2}
\end{align}
From the general formulation given in Appendix \ref{app:A}
for channels with arbitrary length $M$ in the presence of
symmetric rates (\ref{eq:rate}), we obtain by specializing to
the present model with $M=2$,
\begin{align}
\label{eq:j_A1_final-init}
  \left\langle j_{L,1}\right\rangle_t &=
  \left(1-p_1+K\,p_{2,1}^{(3)} \right)\left[ \tilde{k}_{L,1}
  \,e^{\beta\mu_L} - \tilde{k}_{1,L} \, \left(
  \frac{p_{2,1}^{(2)}}{p_{2,1}^{(4)}} \right) \right] ,\\
  \left\langle j_{1,2}\right\rangle_t &=k_{1,2}\, p_{2,1}^{(2)} -
  k_{2,1} \, p_{2,1}^{(3)} \, ,
\label{eq:j_12_final-init} \\
  \left\langle j_{2,R}\right\rangle_t &=
  \left(1-p_2+K\,p_{2,1}^{(2)} \right)\left[ \tilde{k}_{2,1}\,
  \left( \frac{p_{2,1}^{(3)}}{p_{2,1}^{(4)}} \right)
   -\tilde{k}_{R,2} \,e^{\beta\mu_R}
  \right] .
\label{eq:j_2B_final-init}
\end{align}
with $K=\sqrt{\zeta}-1$; $\zeta=e^{-\beta V}$ (for details see
appendix~\ref{app:A}). For the sake of simplified notation, time
arguments on the right hand side of these and subsequent equations
are suppressed. The nearest neighbor correlators $p_{2,1}^{(n)}$
in these equations are defined by Eq.~(\ref{eq:A.30}) with $l=1$.
Note that they are directly related to the average occupations of
system states. Indeed, $P_{11} = p_{2,1}^{(1)}$, $P_{01} =
p_{2,1}^{(2)}$, $P_{10} = p_{2,1}^{(3)}$ and $P_{00} =
p_{2,1}^{(4)}$. Following the Appendix, the two- point correlator
$p_{2,1}^{(1)} \equiv \left\langle n_{2}n_{1}\right\rangle$ is
explicitly found as
\begin{widetext}
\begin{align}
\label{eq:two_point_corr_21}
  p_{2,1}^{\left(1\right)}(t)
  =\frac{1}{2\left(1-\zeta\right)}
  \left[\left(p_{1}+p_{2}\right)\left(1-\zeta\right)-1+\sqrt{\left[
  \left(p_{1}+p_{2}\right)\left(1-\zeta\right)-1\right]^{2}
  +4p_{1}\,p_{2}\zeta\left(1-\zeta\right)}\right]
\end{align}
\end{widetext} Clearly, setting all currents
(\ref{eq:j_A1_final-init}) to (\ref{eq:j_2B_final-init}) equal to
zero implies the equilibrium condition
$\varepsilon_L+\mu_L=\varepsilon_R+\mu_R$ between both reservoirs.
This is seen most directly from the underlying equation
(\ref{eq:A.8}).

In the limit $V \rightarrow 0$ one recovers
$p_{2,1}^{\left(2\right)} \rightarrow p_1\,p_2$. Then
Eqs.~(\ref{eq:j_A1_final-init})-(\ref{eq:j_2B_final-init}) become
\begin{align}
\label{eq:current_A1} \left\langle j_{L,1}\right\rangle_t &=
   k_{L,1}(t)\, p_L\, \left( 1-p_1(t) \right)
   -  k_{1,L}(t)\, p_1(t)\,\left( 1-p_L \right) \\
\label{eq:current_12} \left\langle j_{1,2}\right\rangle_t &=
   k_{1,2}(t)\, p_1(t)\, \left( 1-p_2(t) \right)
   - k_{2,1}(t)\, p_2(t)\,\left( 1-p_1(t) \right) \\
   \left\langle j_{2,R}\right\rangle_t &=  k_{2,R}(t)\,
   p_2(t)\,\left( 1-p_R \right) -  k_{R,2}(t)\, p_R\,\left(
   1-p_2(t)\right) \label{eq:current_2B}
\end{align}
showing that site blocking effects are incorporated in a
mean-field like manner. For general $V$, effective blocking
factors for sites $l=1\, ,2$ can be read directly from
(\ref{eq:j_A1_final-init}) and (\ref{eq:j_2B_final-init}).
Considering site $1$, the effective blocking factor that appears
in the current $\langle j_{L,1} \rangle_t$ is
$1-p_1+K\,p_{2,1}^{(3)}$, which reduces to $1-p_1$ for $V=0$ and
to $1-(p_1+p_2)$ for $V \rightarrow \infty$, because
$p_{2,1}^{(1)}\rightarrow 0$ in the latter case. Hence, for $V
\rightarrow \infty$ the occupation of the nearest neighbor site,
$p_2$, enters into the effective blocking factor additively with
$p_1$. The last conclusion follows from
(\ref{eq:two_point_corr_21}) provided that $p_1+p_2<1$, which
obviously holds for $V \rightarrow \infty$ provided that $\mu_R$
and $\mu_L$ stay finite. \cite{note3} By this, and based on the
identification $P_{11} \equiv p_{2,1}^{(1)}$, it is
straightforward to show that for $V \rightarrow \infty$ the
expressions for the currents
(\ref{eq:j_A1_final-init})-(\ref{eq:j_2B_final-init}) and hence
all results from TDFT become equivalent with those of
Sec.~\ref{sec:4sites} in the same limit (i.e. for $P_{11}
\rightarrow 0$). A similar conclusion holds for $V \rightarrow
-\infty$, where $K_{10,11} \rightarrow \infty$. According to
(\ref{eq:dp_11}) this limit requires that $P_{10} \rightarrow 0$.
Similarly, $P_{01} \rightarrow 0$, and $P_{11}=p_1=p_2=1$
independent of time is seen to be a consistent solution. In this
limit all currents are zero as the channel is jammed by the
presence of two particles. On the other hand,
(\ref{eq:two_point_corr_21}) predicts that $p_{2,1}^{(1)}
\rightarrow \rm max(p_1, p_2)$, which becomes unity since
$p_{1},p_{2} \rightarrow 1$. It follows that the TDFT correctly
yields zero currents in this limit.

The following section provides a quantitative test of TDFT against
the exact results obtained from
Eqs.~(\ref{eq:dp_10})-(\ref{eq:norm}). When applied to the
$4$-site model, this approximate method is not simpler than the
exact treatment, however, its advantage lies in the immediate
applicability to chains of arbitrary length $M$. \cite{Einax}

\section{Results} \label{sec:results}
\subsection{Peristaltic driving, dilute limit}
\label{subsec:resultsA}

\begin{figure}[H!] t
\begin{center}
   \includegraphics[width=0.49 \textwidth]{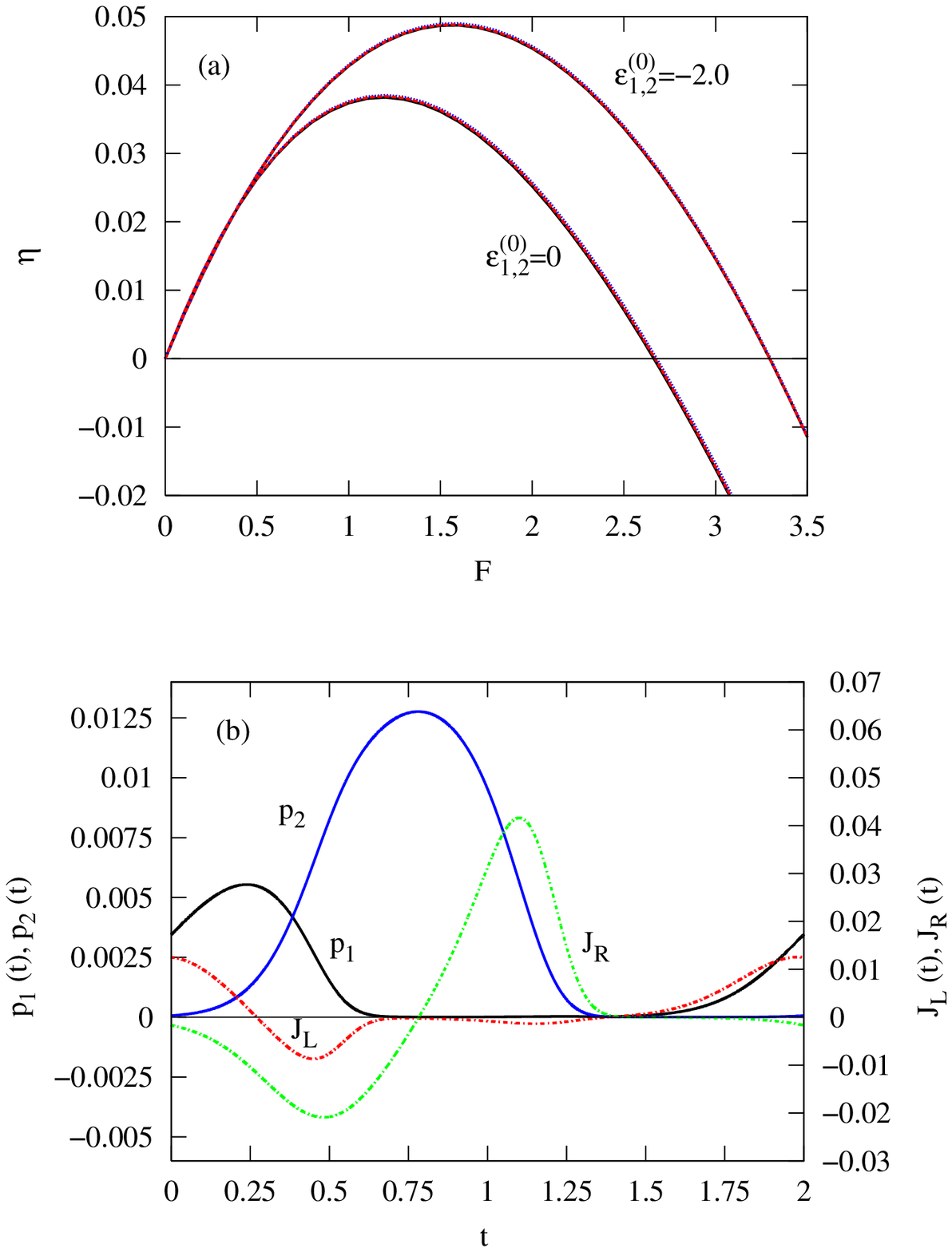}
\end{center}
\caption{Properties of a dilute system with $p_L=p_R=0.0067$
  ($\mu_{L,R}=-5.0$), driving period $\tau=2$ and amplitude $A=5$.
   The behavior of this dilute system does not depend on $V$
   for $V=0, 1$ and $10$, and results from the TDFT approximation
   are indistinguishable from the exact ones. (a) Efficiency $\eta$
   as a function of mechanical bias $F=\varepsilon_R-\varepsilon_L$ ($\varepsilon_L=0$).
   Lower curve: neutral static site energies,
   $\varepsilon_{1,2}^{(0)}= 0$.
   Upper curve: attractive static site
   energies $\varepsilon_{1,2}^{(0)} = -2.0$. (b) Time dependent
  occupation probabilities $p_{1}(t)$, $p_{2}(t)$ and currents $J_{L,R}(t)$ for
  $\varepsilon_{1,2}^{(0)} = -2.0$ and $F=1.5$. Shown is one period of stationary
  oscillation, where $t=0$ corresponds to a minimum in
  $\varepsilon_1(t)$, see Eq.~(\ref{eq:peristaltic_1}).}
\label{fig:1}
\end{figure}

Now we investigate a $1$-D open channel under peristaltic driving
by numerically solving Eqs.~(\ref{eq:dp_10})-(\ref{eq:norm}) in
Sec.~\ref{sec:4sites} and
Eqs.~(\ref{eq:p_1})-(\ref{eq:two_point_corr_21}) in
Sec.~\ref{sec:TDFT} with the input
Eqs.~(\ref{eq:peristaltic})-(\ref{eq:mechanical_gradient}). We
first consider the dilute limit, approached by setting
$\mu_L=\mu_R=-5$, i.e., $p_L=p_R\approx 0.0067$. Calculated
efficiencies as a function of the mechanical load $F$ show a
maximum before they drop to zero and to negative values (current
reversal), see Fig.~\ref{fig:1}. Clearly, for strong dilution,
site blocking effects and the interaction $V$ become irrelevant.
Hence, the efficiency curves in Fig.~\ref{fig:1} become
indistinguishable from the predictions of the independent-particle
model Eqs.~(\ref{eq:p1dt}) and (\ref{eq:p2dt}). Also in this
dilute limit, the TDFT results in Sec.~\ref{sec:TDFT} perfectly
agree with the numerical solutions of the exact equations
(\ref{eq:dp_10})-(\ref{eq:norm}). Indeed, in this limit $P_{11}$
becomes negligibly small, and, as one can easily verify, both the
exact equations (\ref{eq:dp_10})-(\ref{eq:norm}) and the TDFT
equations (\ref{eq:p_1})-(\ref{eq:j_2B_final-init}) become fully
equivalent to the linear rate equations (\ref{eq:p1dt}) and
(\ref{eq:p2dt}).

Figure~\ref{fig:1}a also reveals a significant enhancement of the
efficiency for a channel which is made more attractive relative to
the reservoirs, by changing the static part of site energies in
(\ref{eq:peristaltic_1}) from $\varepsilon_{1,2}^{(0)} =0$ to
$\varepsilon_{1,2}^{(0)} =-2.0$. The value
$\varepsilon_{1,2}^{(0)} =-2.0$ is therefore used in all our
subsequent calculations with peristaltic driving. Apparently,
attracting more particles to the channel interior overcompensates for
the effect of a less favorable exit rate $k_{2,R}$.

Direct insight into the peristaltic mechanism is gained from
Fig.~\ref{fig:1}b. During a time window where $\varepsilon_1(t)
\lesssim 1$ particles from the left reservoir have access to the
channel so that the left current $J_L(t)$ and $p_1(t)$ increase
with time. (This happens for $1.5 \lesssim t \lesssim 2$ in the
plot of Fig.~\ref{fig:1}b.) During the subsequent upward movement
of level $1$ density from this level flows both back to the
reservoir, rendering $J_L(t)<0$, and in the forward direction to
level $2$ , which is lower in energy because of the phase lag
$\pi/2$ between both levels. Therefore $p_2(t)$ increases, but as
long as $\varepsilon_2(t) \lesssim 1$ there is also a flow to
level $2$ from the right reservoir, reflected by the negative peak
in the right current $J_R(t)$. This influx from the right is the
reason why the $p_2$-peak actually gets higher than the
$p_1$-peak. Upward movement of level $2$ in turn causes the
subsequent positive peak in $J_R(t)$, while a decay of $p_2(t)$ to
the left is prohibited as long as
$\varepsilon_2(t)<\varepsilon_1(t)$. The peaks in the
time-dependent densities and currents can be shown to be more
pronounced for $\varepsilon^{(0)}_{1,2}=-2$ in comparison with
$\varepsilon^{(0)}_{1,2}=0$, leading to the increased efficiency
displayed in Fig.~\ref{fig:1}a for the more attractive channel.
Since during the upward movement of level $1$ its population
decays both towards the left and right, we expect that this
mechanism can translocate a particle with probability not larger
than about $0.5$ within one cycle.

\subsection{Peristaltic driving, moderately dense system, $p_L=p_R$}
\label{subsec:resultsB}

\begin{figure}[b]
\begin{center}
    \includegraphics[width=0.49 \textwidth]{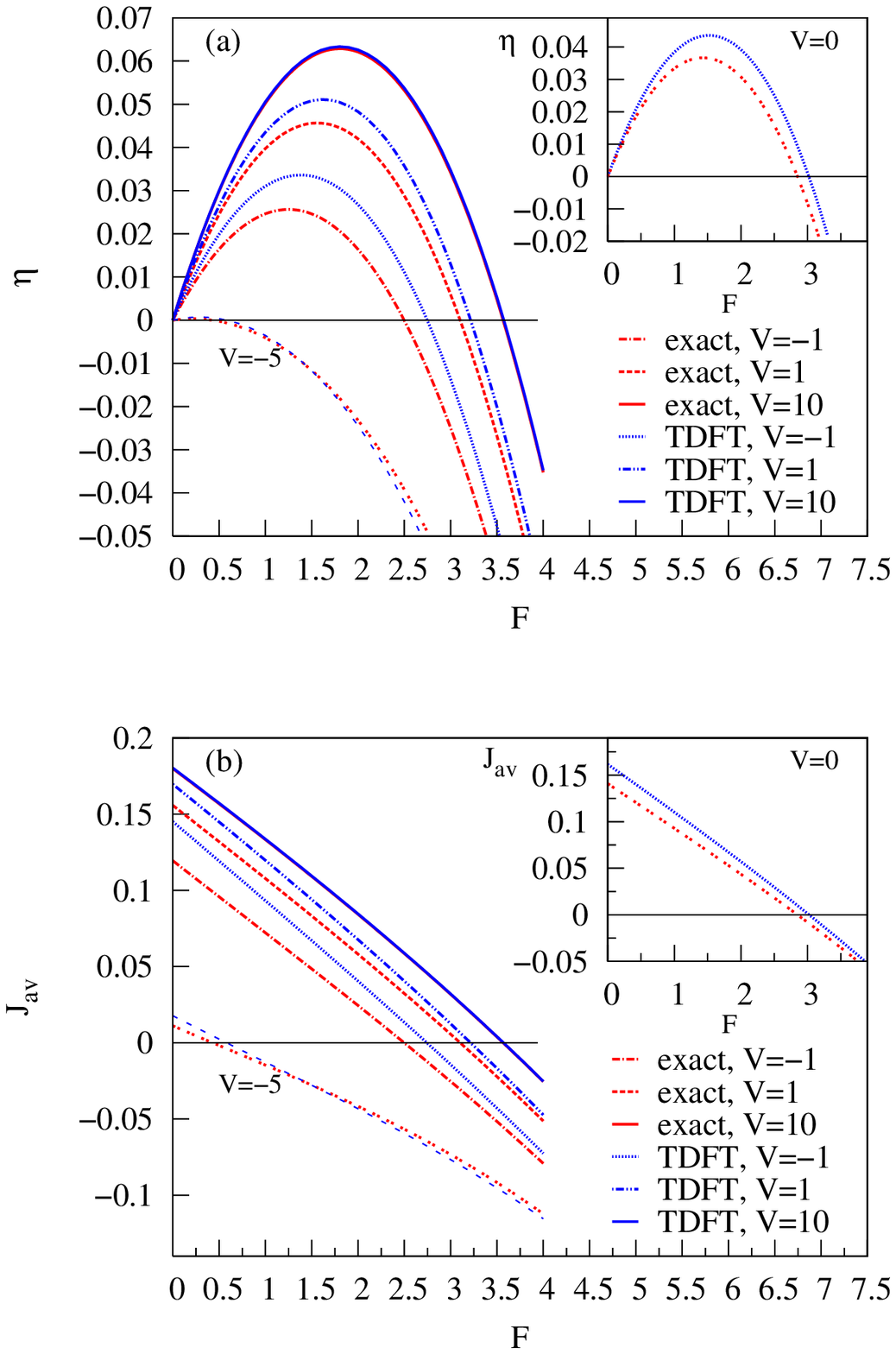}
\end{center}
\caption{(Color online) $F$-dependent efficiency $\eta$ (a) and
  averaged current $J_{\rm av}$ (b) for different interparticle
  interactions $V$ in a driven channel coupled to moderately dense
  reservoirs with $p_L=p_R=0.5$ ($\mu_{L,R}=0$). Other parameters
  are $\varepsilon_{1,2}^{(0)} = -2.0$, $\tau=2$, and $A=5.0$. The
  insets show the $V=0$ case where interparticle interactions arise
  from site exclusion only, and the main figure shows results for
  $V=-5,-1,1,10$. Shown are results from the exact kinetic
  equations (red lines) and from the TDFT (blue lines).} \label{fig:2}
\end{figure}

\begin{figure*}[t]
\begin{center}
   \includegraphics[width=0.99 \textwidth]{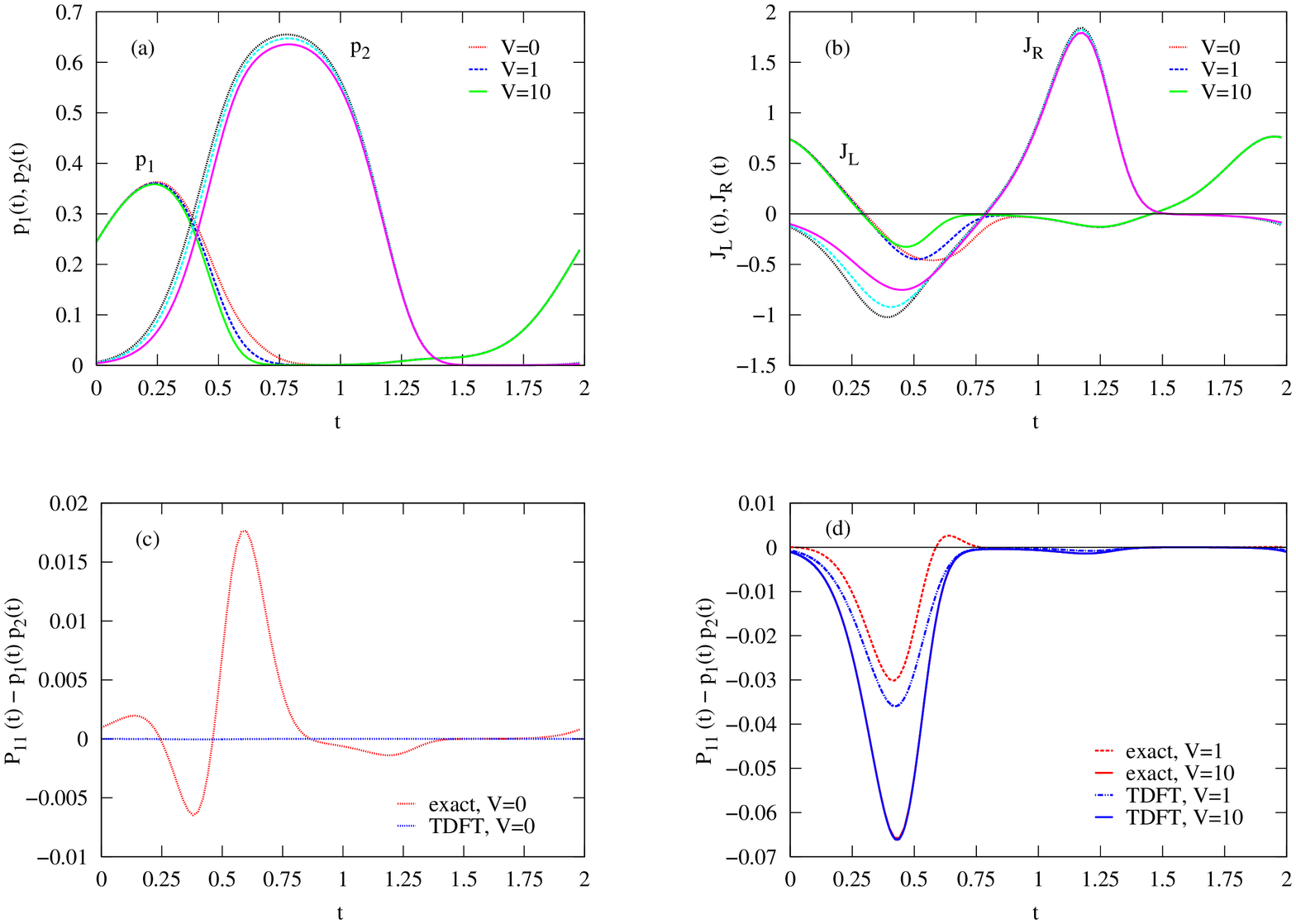}
\end{center}
\caption{(Color online) Exact time dependent occupation
probabilities
  $p_{1,2}(t)$ (a), left and right currents (b), as well as
  correlation function $P_{11}(t) - p_1(t)p_2(t)$ for different $V \ge
  0$ (c and d) in a non-dilute system with $p_{L,R}=0.5$ and $F=1.5$.
  Other parameters are as in Fig.~\ref{fig:2}. The plots (c) and (d)
  illustrate the differences between exact correlations and local
  equilibrium correlations assumed in the TDFT.} \label{fig:3}
\end{figure*}

Figure~\ref{fig:1}b shows that at low densities, a large amplitude
peristaltic oscillation
Eqs.~(\ref{eq:peristaltic})-(\ref{eq:mechanical_gradient}) tends
to locate particles on that site which momentarily has lower
energy. This is evident from the figure as the overlap of the two
peaks for $p_{1}(t)$ and  $p_{2}(t)$ is small. Figs.~\ref{fig:2}
and \ref{fig:3} show results obtained at higher densities, imposed
by setting $p_L=p_{R}=0.5$ ($\mu_L=\mu_R=0$), where effects of
correlations between transporting particles are expected. At these
densities the overlap between the $p_1 (t)$ and $p_2 (t)$ peaks is
larger, as seen in the case $V = 0$ in Figs.~\ref{fig:3}a and
\ref{fig:3}b , which show slightly broader peaks and overlap
regions than in Fig.~\ref{fig:1}b. Consequently, many-body effects
become important, as discussed below.

It may be intuitively expected that simultaneous presence of
particles on both sites $1$ and $2$ should cause reduction of pump
efficiency, because these particles block each other, impairing
the peristaltic driving. Hence one expects that efficiency will
increase as V changes from attractive ($V <0$) to repulsive ($V >
0$), where double occupation of the channel is increasingly
suppressed. These arguments are supported by calculations of
$\eta$ and $J_{{\rm av}}$ for non-dilute systems.
Figure~\ref{fig:2} shows results for $p_{L,R}=0.5$, obtained from
both the exact method described in Sec.~\ref{sec:4sites} and the
TDFT approximation, Sec.~\ref{sec:TDFT}. Before addressing the
quality of the TDFT approximation, let us focus on the main
features in Fig.~\ref{fig:2} common to both treatments.

Comparing the $V=0$ case \cite{note4} shown in the inset to
Fig.~\ref{fig:2}a with the independent particle results (dilute
limit, Fig.~\ref{fig:1}a) we indeed observe that mere site blocking
leads to a decrease in $\eta$. This trend in $\eta$ is enhanced
for increasingly negative $V$ (attractive interparticle
interactions) but is reversed with increasingly positive
(repulsive) $V$, again in agreement with the above expectation.
Interestingly, for $V=10$, $\eta$ significantly exceeds the result
for the dilute limit discussed before, see Fig.~\ref{fig:1}a, by
about $30$ percent. One should note that a strong repulsion as in
the case $V=10$ is practically equivalent to the condition that
the two-site channel can only be singly occupied or vacant. This
``single particle limit'' appears to optimize active transport in
our model.

Similar conclusions hold with respect to the average current
$J_{{\rm av}}$, plotted against $F$ in Fig.~\ref{fig:2}b. The
current increases for increasingly repulsive (positive) $V$ and
becomes negligibly small for increasingly attractive interaction.
For $V=10$ the reversal potential (the load $F$ for which $J_{\rm
av}$ vanishes) is larger by about $15$ percent than the
corresponding value in the case $V=0$. For strong attractive
interaction, e.g. $V=-5$ , see Fig.~\ref{fig:2}, both current and
efficiency are strongly damped.

Regarding the comparison of both methods, it is evident from
Fig.~\ref{fig:2} that the TDFT can well account for important
qualitative features in our model of a driven open channel. At
$V=0$, the TDFT becomes identical to ordinary mean-field theory
because all correlators factorize, leading to
Eqs.~(\ref{eq:current_A1})-(\ref{eq:current_2B}). One should note
that quite generally non-trivial and even long range correlations
can be induced in systems when driven far away from equilibrium
\cite{Spohn83,Derrida07}. Such correlations are ignored in the
TDFT approach. From the inset to Fig.~\ref{fig:2}, we see that for
$V=0$ TDFT overestimates the maximum efficiency $\eta$ by about
$14$ percent relative to the exact calculation. This error shrinks
when $V$ increases from zero, because the TDFT accounts for the
concomitant suppression of double occupation of the channel. For
strong coupling, $V\gg 1$, the TDFT applied to our model
asymptotically becomes exact because, as discussed before in
Sec.~\ref{sec:TDFT}, it correctly describes the one-particle
limit.

For practical applications, attractive interactions are less
interesting than repulsive ones since they tend to immobilize a
pair of particles on sites $1$ and $2$, thus diminishing the
current. For the sake of comparison of the two methods, however,
we have included in Fig.~\ref{fig:2} examples with $V<0$. An
almost quantitative accuracy of the TDFT is observed when the
attraction becomes as strong as $V=-5$, again in accord with the
previous analytical arguments in Sec.~\ref{sec:TDFT}.

The development of correlations inside the channel of the type
discussed above is illustrated in Figs.~\ref{fig:3}c and
\ref{fig:3}d, which show the difference $P_{11}(t)-p_1(t)p_2(t)$
within one driving period. The load is taken as $F=1.5$, where the
corresponding efficiencies are close to their maximum. For $V=0$,
see Fig.~\ref{fig:3}a, peaks of either sign appear in the exact
results in parallel with the peaks in $p_{1}(t)$ and $p_{2}(t)$,
but they are narrower than the latter. On average, these
current-induced nonequilibrium correlations are positive, i.e.,
attractive. However, already under the mild repulsive interaction,
$V=1$, the negative peak takes over. For strong repulsion,
$P_{11}$ practically vanishes, so that the curve for $V=10$ simply
reflects the negative product of the densities $p_{1}$ and
$p_{2}$. Obviously, the TDFT reproduces such correlations better
for larger $V$.

\begin{figure}[t]
\begin{center}
  \includegraphics[width=0.49 \textwidth]{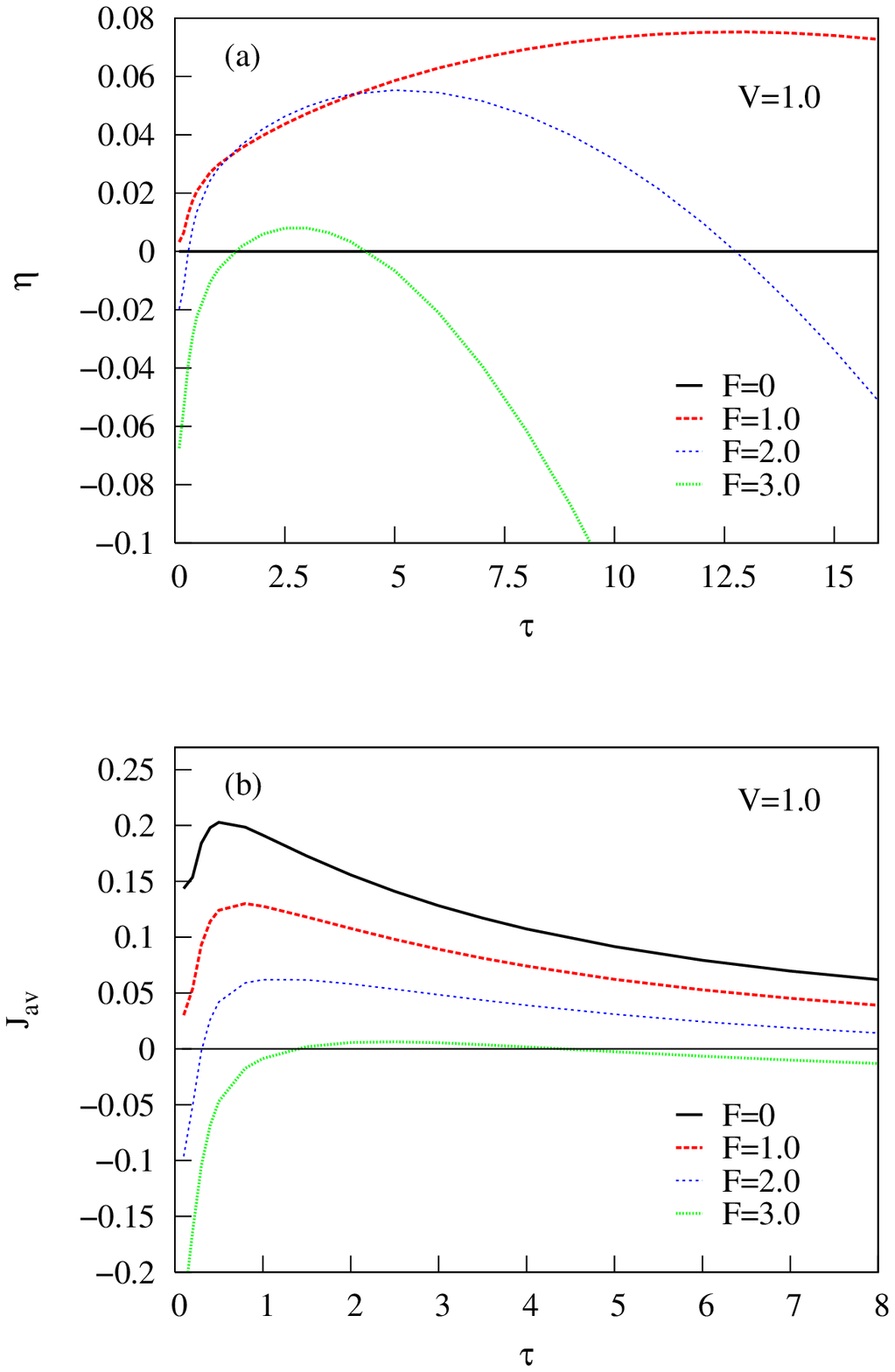}
\end{center}
\caption{(Color online) Efficiency $\eta$ (a) and averaged current
$J_{\rm av}$ (b) from exact rate equations as a function of $\tau$
for
  $p_{L,R}=0.5$, $V=1$ and different $F=0,1,2,3$. Other parameters as
  in Fig.~\ref{fig:2}.} \label{fig:4}
\end{figure}

The influence of the driving period $\tau$ (in fact $\tau/\nu_B$
as discussed above) is displayed in Fig~\ref{fig:4}. Unless
otherwise specified, here and in the following we only show the
exact results from Sec.~\ref{sec:4sites}. In all cases studied,
however, we have verified that the TDFT performs with similar
accuracy as for the foregoing plots. Obviously, except for the
driving mode itself, the driving period is a fundamental parameter
that determines the pump performance. For our model both the the
efficiency $\eta$ and the current $J_{\rm av}$ go through a
maximum as a function of $\tau$. To maximize $\eta$, a smaller
load requires slower driving, as seen in Fig.~\ref{fig:4}a.
Increasing $F$, the maxima in $\eta$ shift to shorter $\tau$. On
the other hand, from the point of view of maximizing $J_{\rm av}$,
even shorter $\tau$ are required, e. g. $\tau \simeq 1.2$ for
$F=2$ (see Fig.~\ref{fig:4}b). This illustrates that optimization
of pump performance in general depends on the optimization
criterion for a particular application. The choice $\tau=2$ in
most of our calculations is found to be a good compromise between
these different optimization criteria. Yet another criterion might
be to maximize the number of particles transmitted within one
cycle, which is $\tau J_{{\rm av}}$. Figure~\ref{fig:4}b suggests
that in our model this number will be limited by about $0.5$, as
argued before. It should also be noted that $\tau$-dependent
efficiencies and the peak structure in the $J_{{\rm av}}$ versus
$\tau$ curves qualitatively agree with
Refs.~\onlinecite{Slanina08,Slanina09}, where a model based on
repulsive on-site interactions and periodic boundary conditions
was considered.

\subsection{Peristaltic pump; effect of chemical driving, $p_L \not=p_R$}
\label{subsec:resultsC}

\begin{figure}[t]
\begin{center}
 \includegraphics[width=0.49 \textwidth]{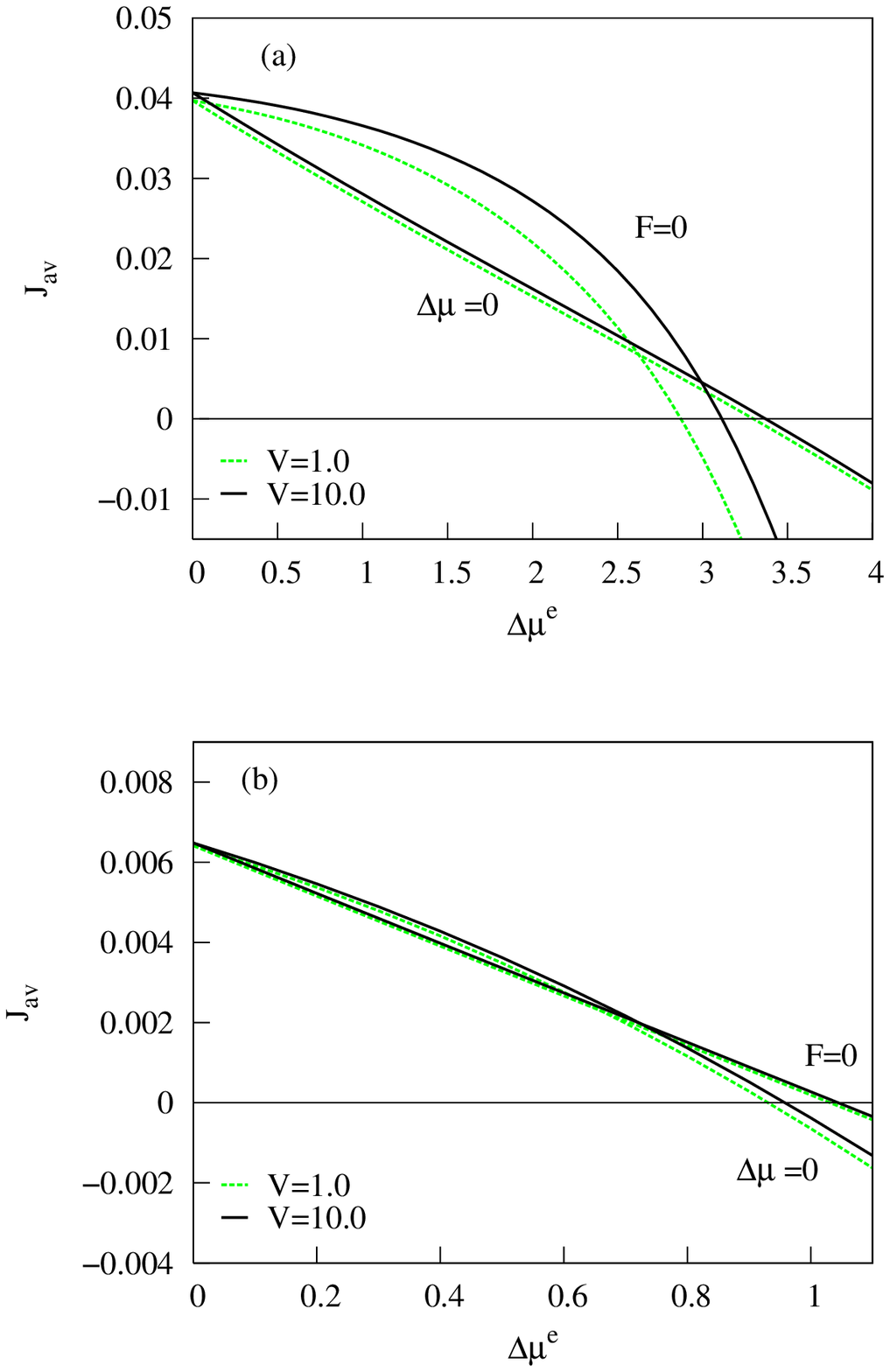}
\end{center}
\caption{(Color online) (a) The average current $J_{\rm av}$ from
  the exact rate equations in a peristaltic pump as a function of the
  electrochemical potential difference $\Delta \mu^{e}= F+ \Delta \mu$
  between the two reservoirs, for $V=1$ and $10$. Other parameters are as in Fig.~\ref{fig:2}:
  $\varepsilon_{1,2}^{(0)}=-2.0$, $\tau=2$, and $A=5.0$. Compared are the two
  limiting cases $J_{\rm av} (F,0)$ where $\Delta \mu=0$ and $J_{\rm av} (0,\Delta \mu)$
  where $F=0$. In either case the left density is fixed to
  $p_L=0.1$ ($\mu_L \approx -2.197$). (b) Same for a flashing ratchet
  with $\varepsilon_{1,2}^{(0)}=-1.0$, $\tau=2$, and $A=2.0$.}
 \label{fig:5}
\end{figure}

It is well-known that passive transport in the linear response
regime at constant temperature is governed solely by the gradient
in the external electrochemical potential
\begin{align}
\label{eq:electrochem_pot}
\mu^{e}&=F+\mu \, .
\end{align}
Far from equilibrium, however, the mechanical and chemical
potential $F$ and $\mu$ play separate roles in determining the
kinetics. This difference stems from the different ways at which
these imposed biases are expressed inside the channel. The
``mechanical bias'' $F=\varepsilon_R-\varepsilon_L$ is assumed to
fall linearly along the channel, as described by
Eqs.~(\ref{eq:peristaltic})-(\ref{eq:flashing_e2}), while the
chemical bias $\Delta \mu = \mu_R-\mu_L$ is assumed to affect only
the external reservoirs $R$ and $L$. Therefore, the average
current $J_{{\rm av}}$ in our model depends differently on these
two external variables, $J_{{\rm av}}=J_{{\rm av}}(F,\Delta\mu)$.
This is illustrated in Fig.~\ref{fig:5}, where the current is
plotted versus the difference $\Delta\mu^{e}=\mu^{e}_R-\mu^{e}_L$
in the electrochemical potentials of the two reservoirs under
different partitionings of $\Delta\mu^{e}$ with respect to the
mechanical and chemical load. In these calculations $p_L=0.1$ is
fixed ($\mu_L \approx -2.197$), whereas $p_R \geq p_L$ is variable
upon varying $\Delta \mu$. Shown are only the two limiting cases
$J_{{\rm av}}(0,\Delta\mu)$ and $J_{{\rm av}}(F,0)$, labelled as
$F=0$ and $\Delta\mu=0$, respectively. Note that in biophysical
systems pumping of charged ions across a membrane will generate a
voltage drop so that in principle the full two-variable
characteristics $J_{{\rm av}}(F,\Delta\mu)$ will enter. As seen
from Fig.~\ref{fig:5}a, the current $J_{{\rm av}}(0,\Delta\mu)$
for small bias is less sensitive to changes in $\Delta\mu$ in
comparison with the $F$-dependence of $J_{{\rm av}}(F,0)$. This is
because, after time averaging, $p_R$ enters the net current from
site $2$ to $R$ mainly through the blocking factor $1-p_R$ in
equation~(\ref{eq:gamma_states}), implying only a mild
$p_R$-dependence. However, when $p_R\gtrsim 0.5$ the current
$J_{{\rm av}}(0,\Delta\mu)$ drops more steeply towards negative
values than $J_{{\rm av}}(F,0)$. In contrast, as in
Fig.~\ref{fig:3}, the drop in $J_{{\rm av}}(F,0)$ down to the
reversal potential is not far from linear. The corresponding TDFT
results were found to be in excellent agreement with these
results. Note that the small concentration $p_L=0.1$ implies that
the interaction $V$ has only a minor influence on $J_{{\rm
av}}(F,0)$.

\begin{figure}[t]
\begin{center}
\includegraphics[width=0.49 \textwidth]{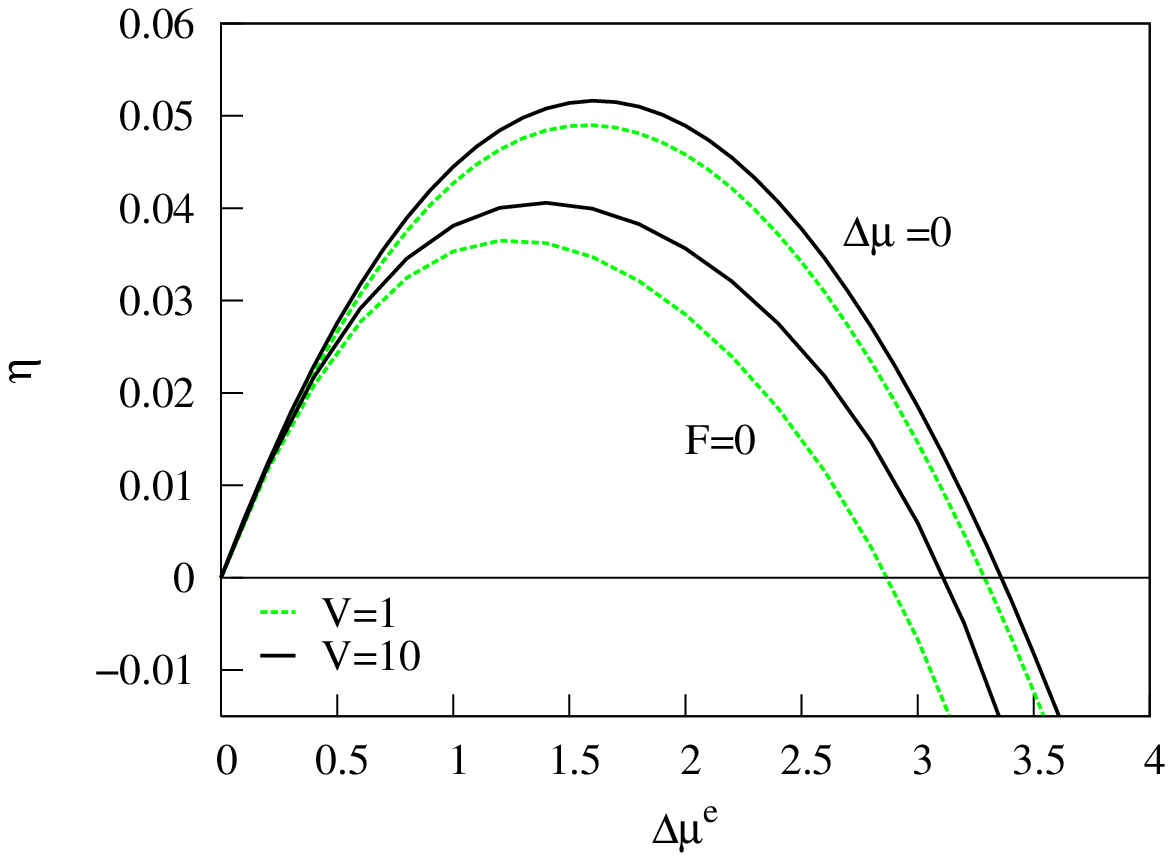}
\end{center}
\caption{(Color online) Chemical efficiency ($F=0$) compared
with mechanical efficiency ($\Delta \mu=0$) for peristaltic
driving, both as a function of $\Delta \mu^{e}=F+ \Delta \mu$
and for different $V$. The left density
is fixed to $p_L=0.1$ ($\mu_L \approx -2.197$).
Other parameters as in Fig.~\ref{fig:5}a.}
 \label{fig:6}
\end{figure}

The difference in the channel's response under a chemical versus
mechanical load also becomes apparent when we compare the
respective efficiencies, which are plotted in Fig.~\ref{fig:6}
against $\Delta \mu^{e}=\mu_R^{e} - \mu_L^{e}$. As for
Fig.~\ref{fig:5} we have taken $p_L=0.1$ fixed and $p_R$ variable.
The chemical efficiency, defined by Eqs.~(\ref{eq:efficiency}) and
(\ref{eq:W_out_chem}), is displayed by the curves labelled $F=0$.
Their maxima and the corresponding chemical reversal potentials
are systematically lower than those referring to the mechanical
efficiency ($\Delta \mu=0$, $F \not= 0$).

\subsection{Flashing ratchet potential}
\label{subsec:resultsD} 
\begin{figure}[t]
\begin{center}
\includegraphics[width=0.49 \textwidth]{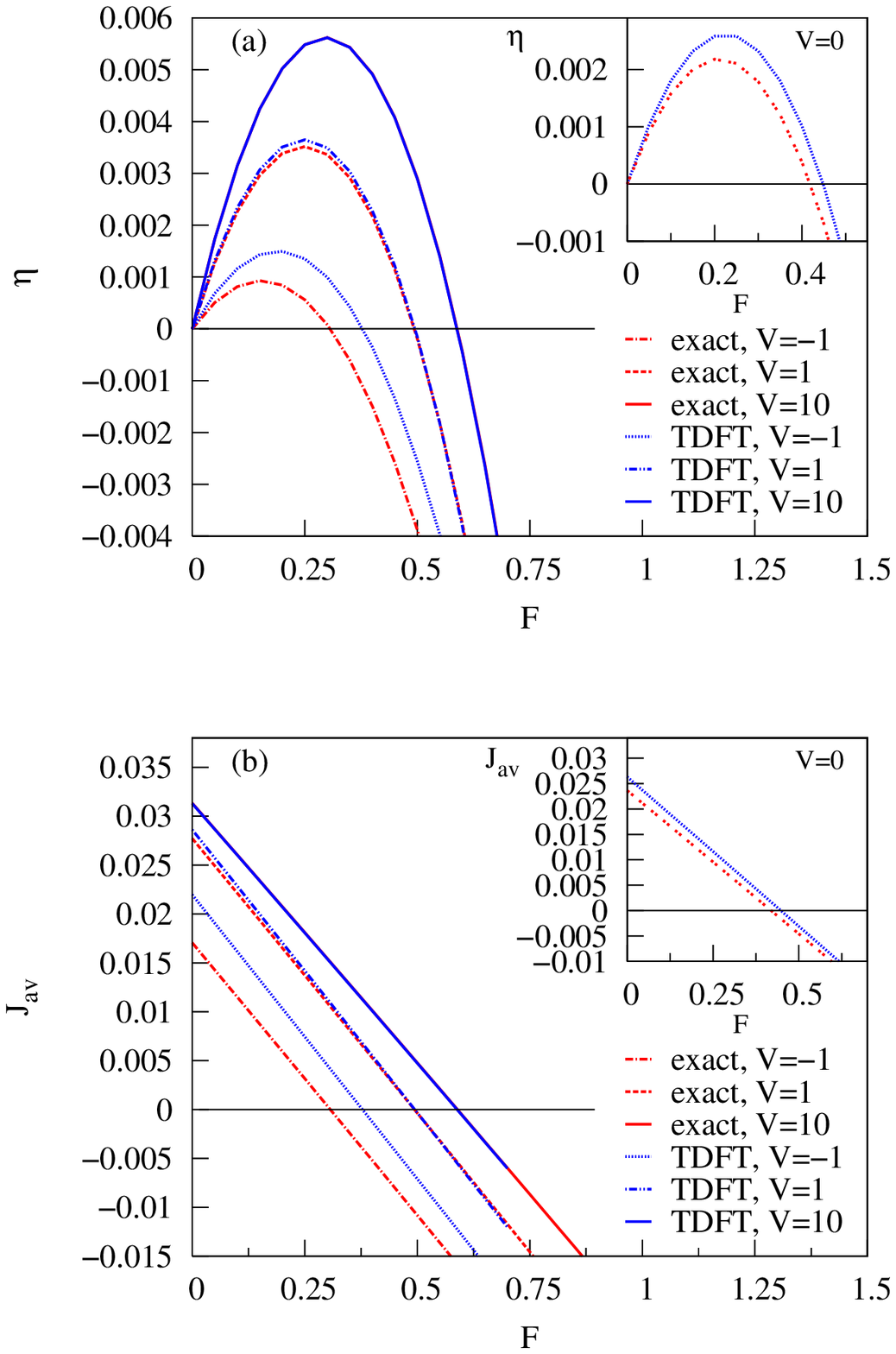}
\end{center}
\caption{(Color online) Efficiency (a) and average current (b),
    plotted against the bias F for a flashing ratchet in the
    moderately dense system characterized by $p_L=p_R=0.5$
    ($\mu_{L,R}=0$). Other parameter are
    $\varepsilon_{1,2}^{(0)} = -1.0$, $\tau=2$, and $A=2.0$.}
 \label{fig:7}
\end{figure}

An important question is to what extent
the previous results for peristaltic driving will change when the
drive mode changes. We exemplify this for the flashing ratchet
potential Eq.~(\ref{eq:flashing}). Here we use the parameters
$A=2.0$ and $\varepsilon_{1,2}^{(0)}=-1.0$ that correspond to a
maximum in $\varepsilon_2(t)$ of the same height relative to the
bath levels as for peristaltic driving with $A=5.0$ and
$\varepsilon_{1,2}^{(0)}=-2.0$. Figure~\ref{fig:7} shows
$F$-dependent mechanical efficiencies $\eta$ and currents $J_{{\rm
av}}$. The qualitative appearance of the curves including their
dependence on $V$ is analogous to Fig.~\ref{fig:2}, but the
absolute performance is considerably less good than for
peristaltic driving. The efficiency $\eta$ and the average current
$J_{\rm av}$ are seen to be lower by about an order of magnitude
and a factor of $5$, respectively, in the flashing driving mode.
Thereby we have verified that choosing $\tau=2$, the efficiency
curves $\eta(F)$ in Fig.~\ref{fig:7} are near optimum. We also
computed the analog to Fig.~\ref{fig:5}a for the flashing ratchet,
see Fig.~\ref{fig:5}b. Currents again are much smaller than in
Fig.~\ref{fig:5}a. A notable feature is the near agreement of the
curves for chemical ($F=0$) and mechanical ($\Delta \mu=0$) bias,
presumably because of a fairly regular distribution of induced
local chemical potential changes along the channel in the case of
the sawtooth potential.

\section{Summary and Conclusions}
\label{sec:Summary}
Active transport was studied within a stochastic model
for an open two-site channel, driven by time-dependent site energies
and coupled to left and right reservoirs. Exact rate equations
for this problem were established and solved numerically. Despite
the simplicity of our model, it allowed us to examine important
conditions for efficient transport, which we expect to be of
relevance regarding both the design of synthetic active channel devices
and, potentially, the functioning of biological motors and ion pumps.

The main findings which emerge from our investigations are
summarized as follows:
\begin{itemize}
\item[{i)}] Because of the coupling to reservoirs with prescribed
   chemical potentials, the number of particles inside the channel
   fluctuates, allowing multiple channel occupation. As a consequence,
   transport depends on interparticle interactions,
   described by a local hard core repulsion and a nearest neighbor
   interaction constant $V$. $V$-dependent motor and pump efficiencies
   are studied systematically, showing that they become optimized
   when multiple occupation of the channel gets suppressed by a
   large repulsive $V$. As intuitively expected, this effect saturates
   and the efficiencies become independent of
   interparticle interaction when $V \rightarrow \infty$.
   In particular, the $V=10$ results displayed in Fig.~\ref{fig:2}
   were found to represent
   this saturation limit. On the other hand, choosing an even
   higher modulation amplitude ($A=10$ instead of $A=5$ used in the
   foregoing calculations for peristaltic driving), the maximum efficiency
   was found to increase only slightly,
   whereas the reversal potential increases considerably.
\item[{ii)}] Different performance criteria can be formulated,
   based, for example, on an economical use of the input energy in producing
   mechanical or chemical work (efficiencies $\eta$), or on
   maximizing the output ``power'', i. e. the current.
   These measures of efficiency, as defined in this
   paper are not intrinsic properties of the system, but obviously depend on
   the imposed load. It is also possible to define an intrinsic measure,
   $\eta_s$, of an ``ideal'' efficiency by (for a mechanical bias;
   a similar definition can be used in the chemical bias case)
   \begin{align}
   \label{eq:eta_s}
   \eta_s &= \frac{ J_{\rm av} (F=0) \, F(J=0) }{ \overline{W}_{\rm in} (F=0)}
   \end{align}
   where $J_{\rm av} (F=0)$ is the flux at zero bias while
   $F(J=0)$ is the mechanical bias for which the current vanishes.
   The product $J_{\rm av} (F=0) \, F(J=0)$ is the analog of the
   product of the short-circuit current and the open-circuit voltage
   that provides an upper bound to the useful
   work that can be extracted from a given voltage source. The choice of
   $\overline{W}_{\rm in} (F=0)$ as the denominator in Eq.~(\ref{eq:eta_s})
   is to some extent arbitrary, since the input work in our
   model is defined through the energy absorbed by the system
   (in contrast to situations encountered, e.g., in photovoltaic
   devices,
   where the input work is defined by the incident rather than
   the absorbed energy).
   Other choices, e.g. using $\overline{W}_{\rm in} (J=0)$
   as denominator in (\ref{eq:eta_s}) could be made.
   As shown in Fig.~\ref{fig:8} these different definitions
   yield different results
   for what may be regarded as the ideal machine performance.
   In either case, the performance is sensitive to the interparticle
   coupling and to the driving period $\tau$,
   and different efficiency criteria require different
   $\tau$ for optimization.
   \begin{figure}[t]
   \begin{center}
   \includegraphics[width=0.49 \textwidth]{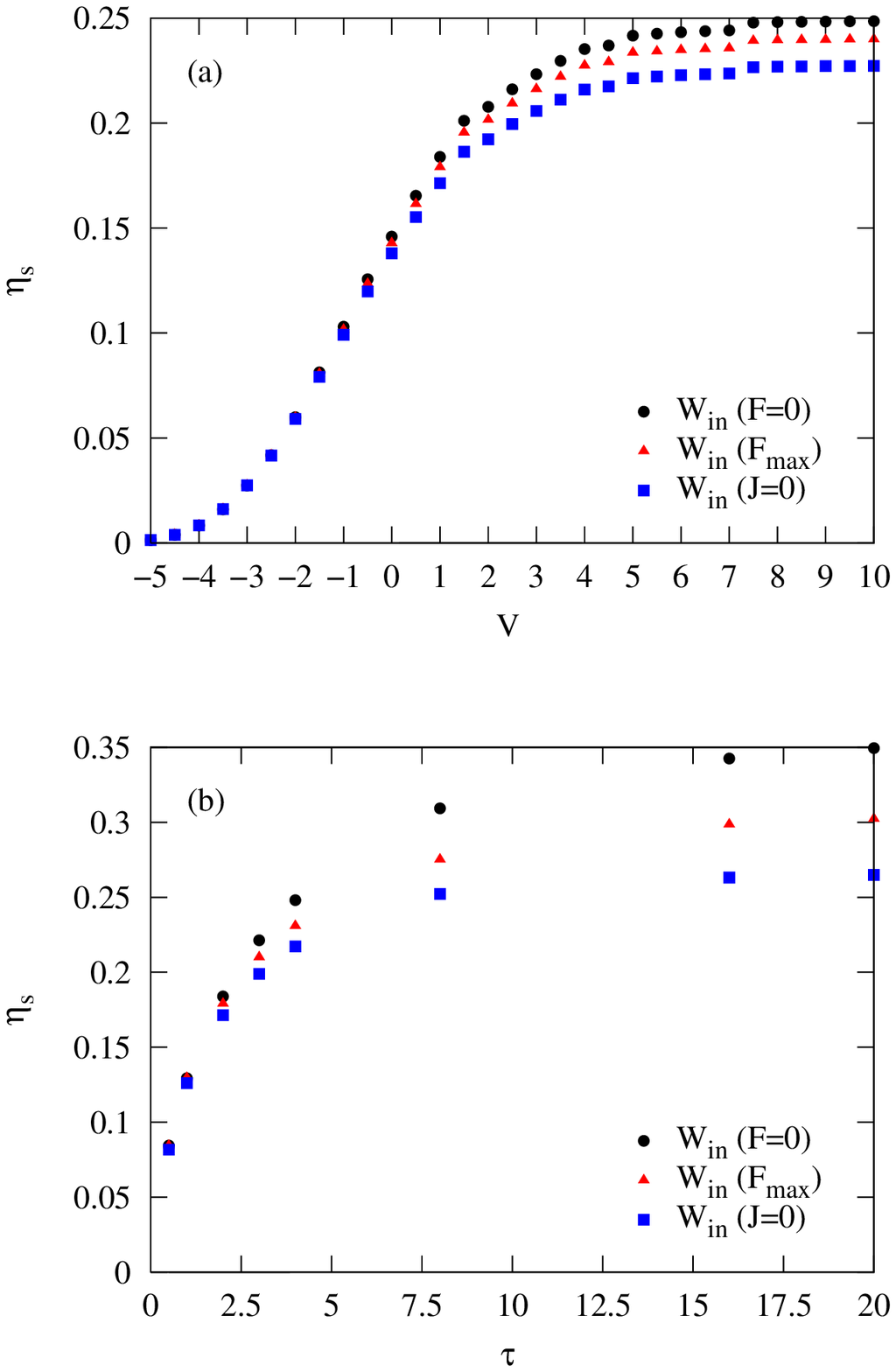}
   \end{center}
   \caption{(Color online) The ideal efficiency $\eta_s$ displayed
   as a function of the interparticle coupling $V$ using
   $\tau=2$ (a) and of the modulation time $\tau$ using $V=1$ (b),
   for a
   peristaltic pump with $\varepsilon_{1,2}^{(0)} = -2.0$, $A=5$
   and three different choices for calculating  $\overline{W}_{\rm in}$:
   at $F=0$, $J=0$ and at $F=F_{\rm max}$, where the efficiency $\eta$
   defined by Eq.~(\ref{eq:efficiency}) is maximal.}
 \label{fig:8}
\end{figure}
\item[{iii)}] In the far-from-equilibrium processes considered here,
   generated currents depend on the partitioning of the electrochemical
   potential difference $\Delta \mu^{(e)}=F+\Delta \mu$ between the reservoirs
   in terms of the mechanical and chemical load $F$ and $\Delta \mu$.
   In other words, these biasing attributes play different roles in
   the kinetics.
\item[{iv)}] Peristaltic driving reflected
   by a phase shift in the oscillating site energies yields
   much better performances than an oscillating sawtooth-like
   potential (flashing ratchet).
\end{itemize}

Some of these findings, in particular the possibility of
efficiency enhancement by repulsive interactions and the
above-mentioned dependencies on the driving period agree on a
qualitative level with earlier work, \cite{Slanina08,Slanina09}
despite considerable differences in the model design and in
details of the results. We take this as an indication that the
corresponding conclusions formulated above should hold rather
generally.

It would certainly be interesting to extend these investigations
to longer channels, with the aim to analyze collective effects in
overdamped systems under time-dependent driving. The full problem
based on the master equation, however, soon becomes intractable as
the length $M$ increases. Results based on the time-dependent
density functional theory, TDFT, nicely agree with the exact
solutions for $M=2$, suggesting that this approximation could
yield reliable results also for arbitrary channel lengths. The
underlying equations of motion for arbitrary $M$ are presented in
the Appendix and will be evaluated in forthcoming work.

It should be emphasized that, unlike equilibrium properties, the
behavior of the systems in non-equilibrium steady states depends,
sometimes sensitively, on the assumed kinetic model. We have
repeated some of the calculations for two more kinetic models, a
normalized kinetic model, where the rates between two levels
with energy separation $\Delta E >0$ are taken to be
\begin{align}
\label{eq:modified_rates1} k {\rm up \choose \rm down} &= \frac{
e^{\mp \frac{\beta}{2} \Delta E} }{e^{\frac{\beta}{2} \Delta E} +
e^{-\frac{\beta}{2} \Delta E}}
\end{align} 
and an asymmetric model where
\begin{align}
\label{eq:modified_rates2} k (\rm up) &= \frac{ e^{-
\frac{\beta}{2} \Delta E} }{1 + e^{-\frac{\beta}{2} \Delta E}} \,
; \quad \quad k (\rm down) = \frac{1}{1 + e^{-\frac{\beta}{2}
\Delta E}} \, .
\end{align} 
Because rates based on Eqs.~(\ref{eq:modified_rates1}) or
(\ref{eq:modified_rates2}) are smaller than those from
Eq.~(\ref{eq:rate}) longer modulation periods are needed in order
to get similar behavior. Apart from this rescaling of $\tau$ the
qualitative behavior of these models remains as with our original
model, Eq.~(\ref{eq:rate}).

Some other important questions are deferred to future work. In
addition to the average currents studied in the present paper,
current fluctuations are observed in experimental studies of
channel transport, and their theoretical evaluation is of
interest. In particular, generalizing available methodologies for
carrier counting statistics to models with time-dependent driving
provides an interesting direction for future investigation.

\begin{acknowledgements}
  ME gratefully acknowledges funding by a
  Forschungsstipendium by the Deutsche Forschungsgemeinschaft (DFG,
  grant number EI 859/1-1). ME and WD have greatly benefited from
  discussions with P. Maass, and WD is also grateful to
  the Max-Planck-Institut fuer Physik komplexer Systeme, Dresden, for its
  hospitality during a stay, where some initial steps of this work
  were carried out. AN acknowledges support by the European Science
  Council (FP7 /ERC grant no. 226628), the Israel - Niedersachsen
  Research Fund, and the Israel Science Foundation. He also thanks
  the AvH Foundation for sponsoring his visit at the University
  of Konstanz. The authors also wish
  to thank the Lion Foundation for supporting this work.
  AN and GS thank Mark Ratner for many helpful discussions.
  This work was also supported by the Non-equilibrium Energy Research
  Center (NERC) which is an Energy Frontier Research Center funded
  by the U.S. Department of Energy, Office of Science, Office of
  Basic Energy Sciences under Award Number DE-SC0000989.
\end{acknowledgements}

\appendix
\section{Currents from the TDFT approach}
\label{app:A} In this Appendix we recall the main steps of the
TDFT for ``Fermionic'' lattice gases
\cite{Reinel96,Kessler02,Gouyet03}. At the same time we generalize
the derivation to systems with arbitrary time dependent site
energies $\varepsilon_{i}$ and kinetic properties governed by a
class of transition rates, where
\begin{align}
  w_{i\rightarrow k}\left(\mathbf{n}\right) &=\nu_{i,k} n_{i} v_{k}
  g\left(H_{i},H_{k}\right)
\label{eq:A.1}
\end{align}
represents the rate for a particle hop from site $i$ to a nearest
neighbor site $k$. Occupation numbers are denoted by $n_{l}$ with $n_{l}=0$ or
$1$, $v_{l}=1-n_{l}$ and $\mathbf{n}\equiv\left\{ n_{l}\right\}$.
$\nu_{i,k}$ are frequency factors. The function $g$ depends on the
energies $H_{i}$ and $H_{k}$ of the particle before and after the
hop, respectively. Focusing on this transition between sites $i$
and $k$, the total lattice gas Hamiltonian is decomposed as
\begin{align}
  H\left(\mathbf{n}\right) &=
  H_{i,k}+n_{i}H_{i}+n_{k}H_{k}+n_{i}n_{k}V_{i,k} \, ,
\label{eq:A.2}
\end{align}
where $V_{i,k}$'s are pair interaction parameters. $H_{i,k}$
denotes the total energy omitting contributions from sites $i$ and
$k$, while
\begin{align}
  H_{i} &=\varepsilon_{i}-\mu_{tot}+\sum_{l\neq
  \left(i,k\right)}V_{i,l}n_{l} \, .
\label{eq:A.3}
\end{align}
Here $\mu_{tot}$ is some reference chemical potential.

Clearly, $H_{i,k}$, $H_{i}$ and $H_{k}$ depend only on
occupational configurations $\hat{\mathbf{n}}$ excluding sites $i$
and $k$. For the following it is convenient to introduce the
quantity
\begin{align}
  G_{i,k}\left(\hat{\mathbf{n}}\right)=g\left(H_{i},H_{k}\right)\exp\left(-\beta
  H_{i}\right) \, .
\label{eq:A.5}
\end{align}
which by the detailed balance condition is required to be
symmetric in its indices,
\begin{align}
  G_{i,k}\left(\hat{\mathbf{n}}\right)&=G_{k,i}\left(\hat{\mathbf{n}}\right)
  \, . \label{eq:A.4}
\end{align}
If transition rates depend only on the particle's initial energy,
 \cite{Kessler02} then $G_{ik}\left(\hat{\mathbf{n}}\right)\equiv
1$. On the other hand, calculations in this work are based on the
symmetric choice $g\left(H_{i}, H_{k}\right) =
\exp\left[\beta\left(H_{i}-H_{k}\right)/2\right]$ or
\begin{align}
   G_{i,k}\left(\hat{\mathbf{n}}\right)&=\exp\left[-\beta\left(H_{i}+H_{k}\right)/2\right]
   \label{eq:A.6}
\end{align}
The TDFT is based on a local equilibrium approximation for the
distribution function,
\begin{widetext}
\begin{align}
  P^{\textrm{loc}}\left(\mathbf{n},t\right)=\frac{1}{Z\left(t\right)}
  \exp\left[-\beta
  H^{\textrm{eff}}\left(\mathbf{n},t\right)\right];\,\,\,
  H^{\textrm{eff}}\left(\mathbf{n},t\right)=H\left(\mathbf{n}\right)+
  \sum_{l}h_{l}\left(t\right)n_{l} \, . \label{eq:A.7}
\end{align}
\end{widetext} Its non-equilibrium character is represented in
terms of time-dependent single-particle fields
$h_{l}\left(t\right)$ that are associated with the time dependent
local chemical potential, see (\ref{eq:A.1}). In view of
Eqs.~(\ref{eq:A.1})-(\ref{eq:A.4}) and using $n_{i}^{2}=n_{i}$;
$n_{i}v_{i}=0$, the average current from $i$ to $k$ can be
expressed as
\begin{widetext}
\begin{align*}
  \left\langle j_{i,k}\right\rangle_{t} & =
  \nu_{i,k}\sum_{\mathbf{n}}\left[ n_{i}v_{k}g\left(H_{i},H_{k}\right)
  - v_{i}n_{k}g\left(H_{k},H_{i}\right)\right]
  P^{\textrm{loc}}\left(\mathbf{n},t\right)\\
  &= \nu_{i,k}\sum_{\mathbf{n}}G_{i,k}\left(\hat{\mathbf{n}}\right)
  \left[n_{i}v_{k}-v_{i}n_{k}\right]
  \exp\left[-\beta\left(h_{i}n_{i}+h_{k}n_{k}\right)\right]
  \exp\left(-\beta H_{i,k}^{\textrm{eff}}\left(\hat{\mathbf{n}}\right)\right)
\end{align*}
\end{widetext} where $H_{i,k}^{\textrm{eff}}$ is to the first term
in (\ref{eq:A.2}) when applied to the corresponding decomposition
of $H^{\textrm{eff}}$. Summation over $n_{i}$ and $n_{k}$ yields
\begin{align}
  \left\langle j_{i,k}\right\rangle_{t} &= M_{i,k}\left(t\right)
  \left[ A_{i}\left(t\right)-A_{k}\left(t\right) \right]
  \label{eq:A.8}
\end{align}
where
\begin{align}
  A_{i}\left(t\right) &=
  \exp\left[-\beta\left(h_{i}\left(t\right)-\mu_{\textrm{eq}}\right)\right]
  \label{eq:A.9}
\end{align}
and
\begin{align*}
  M_{i,k}\left(t\right)=\frac{\nu_{i,k}}{Z\left(t\right)}\sum_{\hat{\mathbf{n}}}G_{i,k}\left(\hat{\mathbf{n}}\right) e^{-\beta
  H_{i,k}^{\textrm{eff}}\left(\hat{\mathbf{n}}\right)} e^{-\beta\mu_{\textrm{eq}}}\,.
\end{align*}
The summation over $\hat{\mathbf{n}}$ can be continued to a
summation over all $\mathbf{n}$ after including vacancy occupation
numbers for sites $i$ and $k$ . Thus
\begin{align}
  M_{i,k}\left(t\right)=\nu_{i,k}\left\langle
  v_{i}v_{k}G_{i,k}\left(\hat{\mathbf{n}}\right)\right\rangle_{t}
  e^{-\beta\mu_{tot}}\, ,
\label{eq:A.10}
\end{align}
which is symmetric, $M_{i,k}(t)=M_{k,i}(t)$, in view
of (\ref{eq:A.5}). Note that $\mu_{tot}$ cancels in
(\ref{eq:A.10}). Density functional theory is employed when
computing average densities $p_{i}\left(t\right)\equiv\left\langle
n_{i}\right\rangle _{t}$ from (\ref{eq:A.7}). The result is the
``structure equation''
\begin{align}
  \varepsilon_{i}+h_{i}+\mu_{i}\left(\mathbf{p}\right)=\mu_{\textrm{tot}}
  \label{eq:A.11}
\end{align}
which allows us to eliminate $h_{i}$ in favor of the local
chemical potential $\mu_{i}\left(\mathbf{p}\right)=\frac{\partial
F}{\partial p_{i}}$; $F\left(\mathbf{p}\right)$ being the free
energy functional associated with $H\left(\mathbf{n}\right)$, with
$\textbf{p}=\{p_l\}$ (for details see
Ref.~\onlinecite{Kessler02}). It follows that
\begin{align}
  A_{i}\left(t\right)=\exp\left[\beta\left(\varepsilon_{i}+
  \mu_{i}\right)\right]\,, \label{eq:A.12}
\end{align}
which can be interpreted as local activity. The essence of
Eq.~(\ref{eq:A.8}) now becomes obvious, namely a factorization of
the average current into a thermodynamic factor (difference of
local activities) and the kinetic coefficient (\ref{eq:A.10}),
which is symmetric, $M_{i,k}=M_{k,i}$. A closed system of
equations determining $p_{i}\left(t\right)$ is obtained by
combining Eqs.~(\ref{eq:A.8})-(\ref{eq:A.12}) with the equation of
continuity
\begin{align}
  \frac{dp_{i}\left(t\right)}{dt}+\sum_{k}\left\langle
  j_{i,k}\right\rangle _{t}=0.  \label{eq:A.continuity}
\end{align}
Now we apply that method to the model of
Sec.~(\ref{sec:Definition}), which is a finite one-dimensional
channel of $M$ sites with nearest neighbor interactions, coupled
to reservoirs $L$ and $R$. From (\ref{eq:A.6}) and (\ref{eq:A.10})
we obtain
\begin{widetext}
\begin{align}
  M_{l,l+1}\left(t\right)&=
  \nu_{l,l+1}\exp\left[-\beta\left(\varepsilon_{l}+
\varepsilon_{l+1}\right)/2\right]
  \left\langle v_{l}v_{l+1}\exp\left[-\beta \left(
  V_{l-1,l}n_{l-1}+V_{l+1,l+2}n_{l+2}\right)
  \right]\right\rangle_{t} \label{eq:A.13}
\end{align}
\end{widetext} where $V_{l-1,l}=V$ for $l=2,...M$. Note that the
reservoirs were taken to have no interaction with the channel,
$V_{0,1} \equiv V_{L,1}=0$ and $V_{M,M+1} \equiv V_{M,R}=0$.
Moreover, it exactly holds that \cite{Kessler02}
\begin{align}
  \beta\mu_{l}\left(\mathbf{p}\right)&=
  \ln\frac{p_{l}}{1-p_{l}}+\ln\frac{\left(1-p_{l}\right)
  p_{l+1,l}^{(2)}}{p_{l}\,p_{l+1,l}^{(4)}}
  +\ln\frac{\left(1-p_{l}\right)p_{l,l-1}^{(3)}}{p_{l}\,p_{l,l-1}^{(4)}}
  \label{eq:A.14}
\end{align}
The correlators $p_{l+1,l}^{(n)}$ are defined by
\begin{align}
\label{eq:A.30}
  p_{l+1,l}^{\left(2\right)} &\equiv \left\langle
  v_{l+1}n_{l}\right\rangle
  = p_{l} - p_{l+1,l}^{\left(1\right)} \\
  p_{l+1,l}^{\left(3\right)} &\equiv \left\langle
  n_{l+1}v_{l}\right\rangle
  = p_{l+1} - p_{l+1,l}^{\left(1\right)}\\
  p_{l+1,l}^{\left(4\right)} &\equiv \left\langle
  v_{l+1}v_{l}\right\rangle =1-p_{l}-p_{l+1}+p_{l+1,l}^{\left(1\right)}
\end{align}
with $n_l=0$ or $1$ the occupation number for site $l$ and
$v_l=1-n_l$ is vacancy occupation number. These quantities
$p_{l+1,l}^{(n)}$ can be determined \cite{Kessler02} from the
``quasichemical condition''
$p_{l+1,l}^{(1)}\,p_{l+1,l}^{(4)}=p_{l+1,l}^{(2)}\,p_{l+1,l}^{(3)}\
e^{-\beta V}$ by solving a quadratic for, e. g., $p_{l+1,l}^{(1)}=
 \left\langle  n_{l+1}n_{l}\right\rangle$.
Equation (\ref{eq:A.14}) is
written as a sum of three terms such that the first term
corresponds to a non-interacting lattice gas, while the second
(third) term is the contribution of the right (left) neighbor to
the chemical potential at site $i$.

Equations (\ref{eq:A.13})-(\ref{eq:A.14}) together with
(\ref{eq:A.12}) determine the currents (\ref{eq:A.8}) in the
interior of the system. Evidently, (\ref{eq:A.13}) involves higher
order correlators, actually up to $4$-point correlators, because
of $\exp\left(\alpha n\right)=1+n\left(\exp\alpha-1\right)$ for
$n=0$ or $1$. Knowing $F(\textbf{p})$, these correlators can be
expressed \cite{Evans79} as functionals of $\textbf{p}$.

The average currents from and to the reservoirs require special
attention. First, because interactions between system and
reservoirs are disregarded, the chemical potentials for the
outermost sites $l=1$ and $l=M$ of the channel satisfy
\begin{align}
  \beta\mu_{1}\left(\mathbf{p}\right)&=
  \ln\frac{p_{2,1}^{(2)}}{p_{2,1}^{(4)}};\,\,\,
  \beta\mu_{M}\left(\mathbf{p}\right)=
  \ln\frac{p_{M,M-1}^{(3)}}{p_{M,M-1}^{(4)}} \label{eq:A.15}
\end{align}
This follows from the fact that all correlators in (\ref{eq:A.14})
involving reservoir sites factorize so that in (\ref{eq:A.14})
contributions of the interactions between sites $l=1, M$ and the
reservoirs vanish. Secondly, the kinetic coefficients that enter
the currents between system and are obtained from (using
Eq.~(\ref{eq:A.13}),
\begin{widetext}
\begin{align}
  M_{L,1}\left(t\right)=\nu_{L}\exp\left[-\beta
  \left(\frac{\varepsilon_{L}+\varepsilon_{1}}{2}\right)\right]
  \left\langle v_{L}\right\rangle \left\langle
  v_{1}\exp\left(-\beta\frac{Vn_{2}}{2}\right)\right\rangle_{t}
\label{eq:A.16},
\end{align}
\end{widetext}
Using again $\exp\left(\alpha n
\right)=1+n\left(\exp\alpha-1\right)$, the last factor in
(\ref{eq:A.16}) simplifies,
\begin{align}
  \left\langle
  v_{1}\exp\left(-\beta\frac{Vn_{2}}{2}\right)\right\rangle
  =1-p_{1}+K\,p_{2,1}^{(3)} \label{eq:A.17}
\end{align}
with $K=\exp\left(-\beta V/2 \right)-1$. An expression analogous
to (\ref{eq:A.16}) is obtained for $M_{M,R}(t)$.

After insertion of (\ref{eq:A.15})-(\ref{eq:A.17}) into
(\ref{eq:A.12}) and (\ref{eq:A.8}) we obtain
\begin{align} \label{eq:j_L}
  \left\langle j_{L,1}\right\rangle_t &=
  \left(1-p_1+K\,p_{2,1}^{(3)} \right)\left[ \tilde{k}_{L,1}
  \,e^{\beta\mu_L}
  - \tilde{k}_{1,L} \, \left(
  \frac{p_{2,1}^{(2)}}{p_{2,1}^{(4)}} \right) \right]\,
\end{align}
where $\left\langle v_{L}\right\rangle$ has been absorbed in the
attempt frequency entering $\tilde{k}_{L,1}$ and $\tilde{k}_{1,L}$
by setting $\tilde{\nu_L}= \left\langle v_{L}\right\rangle \nu_L$.

Similarly,
\begin{align} \label{eq:j_R}
  \left\langle j_{M,R}\right\rangle_t &=
  \left(1-p_M+K\,p_{M,M-1}^{(2)} \right) \left[ \tilde{k}_{M,R} \,
  \left( \frac{p_{M,M-1}^{(3)}}{p_{M,M-1}^{(4)}} \right)
   -\tilde{k}_{R,M} \,\exp(\beta\mu_R)
  \right]\,.
\end{align}
with $\tilde{\nu_R}= \left\langle v_{R}\right\rangle \nu_R$ in the
definition of $\tilde{k}_{R,M}$ and $\tilde{k}_{M,R}$. Note that apart
from modified attempt frequencies $\tilde{\nu}_{L,R}$, the only
reservoir properties entering the theory are the chemical potentials
$\mu_{L,R}$. In this way and by using (\ref{eq:A.continuity}), we end
up with a closed system of equations for an open channel of arbitrary
length, coupled to reservoirs.

Specializing to the $4$-site model ($M=2$) of Sec.~\ref{sec:4sites},
the above expressions for the reservoir currents coincide with
Eqs.~(\ref{eq:j_A1_final-init}) and (\ref{eq:j_2B_final-init}) in the
main text. Moreover, to get the current $\left\langle
  j_{1,2}\right\rangle_t$ inside the two-site channel, we use
(\ref{eq:A.13}) to obtain
\begin{align}
  M_{1,2}\left(t\right)&=\nu\exp\left[
  -\beta\left(\frac{\left(\varepsilon_{1}+\varepsilon_{2}\right)}{2}\right)\right]\left\langle
  v_{1}v_{2}\right\rangle_{t} \label{eq:A.18}
\end{align}
with $\left\langle v_{1}v_{2}\right\rangle_{t}\equiv
p_{2,1}^{(4)}$. Together with (\ref{eq:A.15}), this leads to
(\ref{eq:j_12_final-init}).


\begin{thebibliography}{99}
\bibitem{Juelicher97} F. J\"{u}licher, A. Ajdari, and J. Prost,
        Rev. Mod. Phys. \textbf{69}, 1269 (1997).

\bibitem{Lauger91} P. L\"{a}uger, \textit{Electrogenic Ion Pumps}
        (Sinauer Associates, Sunderland, MA, 1991).

\bibitem{Hille01} B. Hille, \textit{Ionic channels of excitable membranes}
        (Third edition, Sinauer Associates, Sunderland, MA, 2001).

\bibitem{Muneyuki00} E. Muneyuki and T. A. Fukami,
        Biophys. J. \textbf{78}, 1166 (2000).

\bibitem{Kaila08} V. Kaila, M. Verkhovsky, G. Hummer, and M. Wikstr\"{o}m,
        Proc. Natl. Acad. Sci. \textbf{105}, 6255 (2008).

\bibitem{Rousselet94} J. Rousselet, L. Salome, A. Ajdari and J. Prost,
        Nature \textbf{370}, 446 (1994).

\bibitem{Siwy02} Z. Siwy and A. Fuli\'{n}ski,
        Phys. Rev. Lett. \textbf{89}, 198103 (2002).

\bibitem{Galperin05} M. Galperin, M. A. Ratner and A. Nitzan,
        Nanoletters \textbf{5}, 125 (2005).

\bibitem{Altshuler99} B. L. Altshuler and L. I. Glazman,
        Science \textbf{283}, 1864 (1999).

\bibitem{Reimann02} P. Reimann, Phys. Rep. \textbf{361}, 57 (2002).

\bibitem{Haenggi09} P. Haenggi and F. Marchsoni,
        Rev. Mod. Phys. \textbf{81}, 387 (2009).

\bibitem{Derenyi95} I. Derenyi and T. Vicsek,
        Phys. Rev. Lett. \textbf{75}, 374 (1995).

\bibitem{Savelev05} S. Savel'ev, F. Marchesoni and F. Nori,
        Phys. Rev. E \textbf{71}, 011107 (2005).

\bibitem{Reimann99} P. Reimann, R. Kawai, C. Van den Broeck and P. Haenggi,
        Europhys. Lett. \textbf{45}, 545 (1999).

\bibitem{Slanina08} F. Slanina, Europhys. Lett. \textbf{84}, 50009 (2008).

\bibitem{Slanina09} F. Slanina, J. Stat. Phys. \textbf{135}, 935 (2009).

\bibitem{Reinel96} D. Reinel and W. Dieterich,
        J. Chem. Phys. \textbf{104} 5234 (1996).

\bibitem{Kessler02} M. Kessler, W. Dieterich, H. L. Frisch, J. F. Gouyet,
        and P. Maass, Phys. Rev. E \textbf{65}, 066112 (2002).

\bibitem{Gouyet03} J. F. Gouyet, M. Plapp, W. Dieterich and P. Maass,
        Advances in Physics \textbf{52}, 523 (2003).

\bibitem{note1} In the continuum limit, and under the
        assumption of symmetric hopping rates this method has been shown
        \cite{Gouyet03} to become equivalent with dynamic density
        functional theory \cite{Marconi99,Loewen09} used in soft matter
        physics, see also \cite{Dieterich90}.

\bibitem{Dieterich90} W. Dieterich, H. L. Frisch and A. Majhofer,
        Z. Phys. B \textbf{78}, 317 (1990).

\bibitem{Marconi99} U. Marini Bettolo Marconi and P. Tarazona,
        J. Chem. Phys. \textbf{110} 8032 (1999).

\bibitem{Loewen09} P. Espa\~{n}ol and H. L\"{o}wen, J. Chem. Phys.
        \textbf{131}, 244101 (2009).

\bibitem{Kolomeisky07} A. B. Kolomeisky, Phys. Rev. Lett. \textbf{98},
        048105 (2007).

\bibitem{Berezhkovskii/etal:2008} A. M. Berezhkovskii and S. M. Bezrukov, Phys. Rev. Lett. \textbf{100},
        038104 (2008).

\bibitem{Zilman/etal:2009} A. Zilman, J. Pearson, and G. Bel, Phys. Rev. Lett. \textbf{103},
        128103 (2009).

\bibitem{Einax09} M. Einax, M. K\"{o}rner, P. Maass and A. Nitzan,
    Phys. Chem. Chem. Phys. \textbf{12}, 645 (2010).

\bibitem{Dierl/etal:2010} M. Dierl, P. Maass, and M. Einax,
        in preparation.

\bibitem{Derrida:1998} B.\ Derrida, Phys.\ Rep., 1998, \textbf{301}, 65 (1998).

\bibitem{Schuetz:2001} G.\ Sch\"utz, in {\it Phase Transitions in
     Critical Phenomena}, edited by C.\ Domb and J.\ Lebowitz (Academic
     Press, San Diego, 2001), Vol.\ 19, pp.\ 3-251

\bibitem{Sumithra01} K.\ Sumithra and T.\ Sintes
    Physica A \textbf{297}, 1 (2001).

\bibitem{Percus94} J. K. Percus, Acc. Chem. Res. \textbf{27}, 8 (1994)

\bibitem{Buschle00} J. Buschle, P. Maass and W. Dieterich,
    J. Physics A \textbf{33}, L41 (2000).

\bibitem{note3} If $p_1+p_2 >1$, a situation that can be reached
    if $\mu_L$ and/or
    $\mu_R$ approach infinity at finite interaction $V$, a different
    solution is obtained as $V\rightarrow \infty$ that is not
    considered here.

\bibitem{Einax} M. Einax, in preparation

\bibitem{Evans79} R. Evans, Adv. Phys. \textbf{28}, 143 (1979).

\bibitem{note4} Since the numerator and denominator in
  Eq.~(\ref{eq:two_point_corr_21}) become zero for $\zeta=0$, the
  TDFT results labeled $V=0$ actually are for $V=0.001$.

\bibitem{Spohn83} H. Spohn, J. Phys. A: Math. Gen. \textbf{16}, 4275 (1983).

\bibitem{Derrida07} B. Derrida, J. Stat. Mech. P07023 (2007).

\end{thebibliography}
\end{document}